\documentclass{article}

\usepackage{algorithm}
\usepackage{algpseudocode}
\usepackage{amsfonts}
\usepackage{amsmath}
\usepackage{graphicx}
\usepackage{url}
\usepackage[colorlinks=true, urlcolor=blue, linkcolor=red]{hyperref}

\DeclareMathOperator{\tr}{tr}

\begin{document}

\title{Fast Approximate Determinants \\ Using Rational Functions}

\author{Thomas Colthurst\thanks{Google, Inc.  Correspondence may be emailed to {\tt thomaswc@google.com}.},
Srinivas Vasudevan\footnotemark[1], \\
James Lottes\footnotemark[1], and
Brian Patton\footnotemark[1]}

\maketitle

\begin{abstract}
We show how rational function approximations to the logarithm, such as
$\log z \approx (z^2 - 1)/(z^2 + 6z + 1)$,
can be turned into fast algorithms for approximating the determinant
of a very large matrix.  We empirically demonstrate that when combined
with a good preconditioner, the third order rational function approximation
offers a very good trade-off between speed and accuracy when measured on
matrices coming
from Matérn-$5/2$ and radial basis function Gaussian process kernels.
In particular, it is significantly more accurate on those matrices than the
state-of-the-art stochastic Lanczos quadrature method for approximating determinants while running at about the same speed.
\end{abstract}

\section{Introduction}

The problem of calculating the determinant of a large matrix comes up in
numerous fields, including physics \cite{lee2003zone} and
geo-statistics \cite{PACE2004179}.
We are particularly motivated by its application in
the training of Gaussian processes \cite{ebden2015gaussian}, a popular
statistical model for doing non-parameteric inference.

Strassen proved that the computational
complexity of calculating the determinant is the same as that of matrix
multiplication \cite{strassen1969gaussian}, \cite{Strassen1973}, for which the
best practical algorithms for a $n$ by $n$ matrix are O($n^{2.807\ldots}$)
\cite{strassen1969gaussian}.
For large matrices, with $n > 10^6$, these are prohibitively slow, so we
are forced to consider approximate algorithms, such as those presented in
\cite{boutsidis2017randomized}, \cite{zhang2007approximate},
\cite{han2015large}, and \cite{ubaru2017fast}.  All of these approximate
algorithms have time complexity $O(n^2)$, as does the one we present
below.

Our approach to estimating determinants is based on the following
approximations to $\log(z)$ introduced by \cite{kelisky1968rational}:

\begin{eqnarray}
\label{ratdefs}
r_1(z) & = & 2 \frac{z - 1}{z + 1} \notag \\
r_2(z) & = & 4 \frac{z^2 - 1}{z^2 + 6z + 1} \notag \\
r_3(z) & = & \frac{2}{3} \frac{7z^3 + 27z^2 - 27z - 7}{z^3 + 15z^2 + 15z + 1} \notag \\
r_4(z) & = & \frac{16}{3} \frac{z^4 + 10 z^3 - 10z - 1}{z^4 + 28z^3 + 70z^2 + 28z + 1} \notag \\
r_5(z) & = & \frac{2}{15} \frac{43 z^5 + 825 z^4 + 1150 z^3 - 1150 z^2 - 825 z - 43}{z^5 + 45z^4 + 210z^3 + 210z^2 + 45z + 1} \notag \\
r_6(z) & = & \frac{4}{15} \frac{23 z^6 + 708 z^5 + 2355 z^4 - 2355 z^2 - 708 z - 23}{z^6 + 66 z^5 + 495 z^4 + 924 z^3 + 495 z^2 + 66 z^1 + 1} \notag \\
\end{eqnarray}

These approximations were chosen as to minimize $|z r'(z) - 1|$
over a real interval, subject to $r(1) = 0$, and have several nice properties including $r_n (z) = - r_n(1/z)$.
A graph of their respective approximation errors to $\log z$ is shown in
figure~1.  Because the even order approximations $r_2, r_4, \ldots$ offer only
small incremental improvements in accuracy, the rest of this paper focuses
on the odd order approximations $r_1$, $r_3$, and $r_5$.

\begin{figure}[!b]
	\includegraphics[width=11cm]{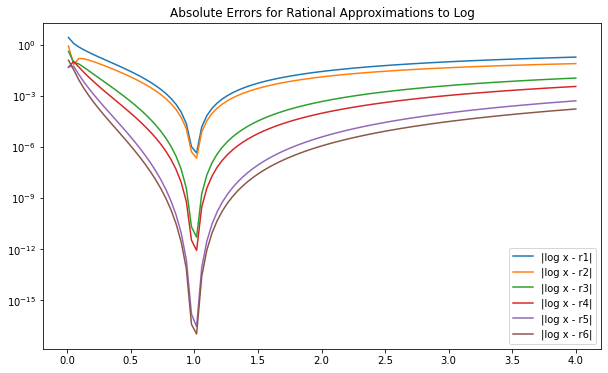}
	\caption{$\log z - r_i(z)$ for the log approximations defined in equation~1.}
        \label{reterrors}
\end{figure}

These rational functions can be directly applied to matrices to obtain
approximations to the matrix logarithm \cite{gimeno2017computation}.
For example,
$$r_3(M) = \frac{2}{3} (7M^3 + 27M^2 - 27M - 7)(M^3 + 15M^2 + 15M + 1)^{-1}$$
is a decent approximation to
$\log M$ for matrices near the identity.  We can then use the identity
$$\log \det M = \tr \log M$$
along with Hutchinson's trick \cite{hutchinson}
$$\tr A = E[z^t A z]$$
for "probe" vectors $z$ of i.i.d. random variables with mean~0
and variance~1 to get the approximation
$$\log \det M \approx (1/n) \sum_{i = 1}^{n} {z_i}^t r(M) z_i$$
This forms the heart of the method that we present in the next section.

\section{Method}

The rational functions introduced in the previous section
are only accurate approximations to $\log$ on scalar values near $1$, so
when extending their domain to matrices, they will be inaccurate if their
input has eigenvalues far from~1.  Also, as pointed out by
\cite{gimeno2017computation}, even if a scalar approximation is accurate
for all of the eigenvalues of a matrix, it can still be inaccurate when
lifted into a matrix function if the input matrix has a high condition
number.

For both of these reasons, we are motivated to combine our
$\log$ approximations with a preconditioner -- an easy-to-compute approximation
$P$ to the matrix $M$ with the properties that
\begin{itemize}
	\item The matrix-vector products $Pv$ and $P^{-1}v$ are easy to compute,
		preferably in time closer to $O(n)$ than $O(n^2)$,
	\item $\det P$ is easy to compute, and
	\item $M P^{-1}$ is closer to the identity than $M$, and in particular
		has a lower condition number.
\end{itemize}
Given such a preconditioner $P$, we can calculate $\log \det M$ as
$$\log \det M = \log \det M P^{-1} P = \log \det P + \log \det M P^{-1} = \log \det P + \tr \log M P^{-1}$$
Beyond the reasons already mentioned, the preconditioner will also help by
lowering the variance of the trace estimate, as that variance is governed by
matrix norms like the Frobenius norm $||M||_F^2 = \sum_{i, j} |M_{ij}|^2$
which a good preconditioner will also tend to decrease.
\cite{martinsson2020randomized}

The primary preconditioner we use in this work is a randomized, truncated
SVD.  This preconditioner was chosen based on the analysis in
\cite{wenger2022preconditioning} which shows that such preconditioners
of size $l$ typically reduce the Frobenius norm of $M - P$
by a factor of $l^{-1/2}$.
Specifically, we use a simplified randomized SVD scheme based on Algorithm 5.3
in \cite{halko2011finding}, where we compute a smaller randomized
orthonormal matrix that approximates the range of $M$ (via Algorithm 4.1), and
construct an SVD based around this smaller matrix.

Given a preconditioner, our $\log \det$ algorithm is presented in
Algorithm~1.  
It depends on the rational functions $r_k$ from
the previous section being written in partial fraction form; those forms
are given in table 1.  
As proved in \cite{kelisky1968rational} the
denominators of the $r_k$ always have negative real roots
(and thus the partial fraction denominators do as well).

\vspace{5mm}

\fbox{\begin{minipage}{30em}
	\label{mainalg}
	\vspace{1mm}
	{\bf Algorithm 1} The r* algorithm for approximating $\log \det m$ \\
	\vspace{-1mm}
	\hrule
	\vspace{2mm}
	{\bf Inputs:}
	\begin{itemize}
			\item An $n$ by $n$ symmetric positive definite matrix $M$,
			\item A preconditioner $P$ of $M$,
			\item A rational approximation $r_k$ to $\log x$
				given as a partial fraction\\
				$r_k(x) = b + \sum_j \frac{c_j}{x - \alpha_j}$,
			\item A mean $0$, variance $1$ distribution $D$ on $\mathbb{R}$, and
			\item Positive integers $s$ and $t$.
	\end{itemize}
        {\bf Start}
	\begin{enumerate}
		\item Create $s$ probe vectors $\{v_i \in \mathbb{R}^n\}$
		       with entries sampled from $D$.
	       \item Run the Lanczos algorithm \cite{lanczos1950iteration}
		for $t$ iterations on $M P^{-1}$ to
	get a $t$ by $n$ matrix $Q$ and an $t$ by $t$ tridiagonal
	matrix $T$ such that
			$$M P^{-1} \approx Q^t T Q$$
\item For each $\alpha_j$ and each probe vector $v_i$, solve the tridiagonal
	system
			$$(T - \alpha_i) w_{i, j} = |v_i| e_1$$
      for $w_{i, j}$.
	\end{enumerate}
	{\bf End} \\
	{\bf Output:}
	$$\log \det P + \frac{1}{s} \sum_i {v_i}^t \left(b v_i +  \sum_j c_j Q^t w_j \right)$$
\end{minipage}}

\vspace{5mm}

\begin{table}
	\centering
\begin{tabular}{lc}
	$r_1(x) = $ & $2 - \frac{4}{x+1}$ \\
	& \\
	$r_3(x) = $ & $\frac{14}{3} - \frac{49.52250037431294}{x+13.92820323027551} - \frac{20/9}{x+1} - \frac{0.2552774034648563}{x+0.0717967697244908} $ \\
	& \\
	$r_5(x) = $ & $\frac{86}{15} - \frac{140.08241129102026}{x+39.863458189061411} - \frac{6.1858406006156228}{x+3.8518399963191827}$ \\
	\rule{0pt}{4ex} & $- \frac{92/75}{x+1} - \frac{0.41692913805732562}{x+0.25961618368249978} - \frac{0.088152303639431204}{x+0.025085630936916615}$ \\
	& \\
\end{tabular}
	\caption{The partial fraction decompositions of the rational approximations to $\log x$ used in this paper.  All decimal values are approximated to 15 places.}
	\label{partials}
\end{table}

Steps~2 and 3 of Algorithm~1 
are in effect a multi-shift solver \cite{jegerlehner1996krylov}, which
solves the equations
$$(A + \sigma_j I) x_j = b$$
for a variety of $\sigma_j$ values by approximating $A$ as $Q^t T Q$
and then manipulating it as
\begin{eqnarray*}
	(Q^t T Q + \sigma_j I) x_j & = & b \\
	(T Q + \sigma_j Q) x_j & = & Q b \\
	(T Q + \sigma_j Q) x_j & = & |b| e_1 \\
	(T + \sigma_j I) Q x_j & = & |b| e_1 \\
	x_j & = & |b| Q^t (T + \sigma_j I)^{-1} e_1 \\
\end{eqnarray*}
with $Q b = |b| e_1$ coming from the fact that $b$ is fed into the Lanczos
algorithm as the initial direction for the construction of $Q$.

With that information, we can now justify the algorithm's output as the
approximation

\begin{eqnarray*}
\log \det M & = & \log \det P + \log \det M P^{-1} \\
            & = & \log \det P + \tr \log M P^{-1}  \\
	    & \approx & \log \det P + \tr \mbox{r}(M P^{-1}) \\
	    & \approx & \log \det P + \tr \left(b I  + \sum_j (c_j (M P^{-1} - \alpha_j I)^{-1} \right) \\
	    & \approx & \log \det P + (1/s) \sum_i {v_i}^t \left(b I + \sum_j c_j (M P^{-1} - \alpha_j I)^{-1} \right) {v_i} \\
	    & \approx & \log \det P + (1/s) \sum_i {v_i}^t \left(b I + \sum_j c_j (Q^t T Q - \alpha_j I)^{-1} \right) {v_i} \\
	    & \approx & \log \det P + (1/s) \sum_i {v_i}^t \left(b v_{i} + \sum_j c_j Q^t w_{i, j} \right) \\
\end{eqnarray*}

It is important to leave the $b I$ terms inside the trace estimate when using
Gaussian probe vectors with entries from $N(0, 1)$.  This is because the
variance of the trace estimate in that case is governed by the Frobenius
norm \cite{epperly2023},
and the positive $b I$ reduces the Frobenius norm given that all of
the $c_j$'s are negative.  However, it makes no difference when using Rademacher
probe vectors (which have values $+1$ and $-1$ each with probability $1/2$)
because there the variance is a function of the off-diagonal entries
\cite{epperly2023}, and adding $b I$ does not affect those.

In terms of time complexity, step 2 of the algorithm takes time $O(t n^2)$ and step 3 takes time $O(s t n)$ since each tridiagonal solve can be done in time $O(n)$ using the Thomas algorithm \cite{thomas1949elliptic}.

\section{Results}

We have implemented Algorithm~1 as part of the open source
Tensorflow Probability package \cite{dillon2017tensorflow} available at \\
\href{https://github.com/tensorflow/probability/tree/main/tensorflow\_probability/python/experimental/fastgp/fast\_log\_det.py}{
{\tt https://github.com/tensorflow/probability/tree/main/tensorflow\_} \\
{\tt probability/python/experimental/fastgp/fast\_log\_det.py}}.
Along with Algorithm 1, we have also implemented the stochastic Lanczos
quadrature (SLQ) algorithm from \cite{ubaru2017fast}
and the conjugate gradients based algorithm
for the gradient of $\log \det$ from \cite{wang2019exact}.
All of this code is implemented in JAX \cite{jax2018github}.

Using this implementation, we then measured the speed and accuracy of the
r* and SLQ algorithms on the covariance matrices generated by the
Matérn-$5/2$ and radial basis function (RBF) Gaussian process kernels
\cite{williams2006gaussian}.  Both kernels had their amplitude and
length scale set to $1$, and used index points sampled from a normal
distribution.
The graph of the measurements, as a function of the matrix's size $n$, are
plotted in
figures~2 through 5.  The speeds include the time required to caculate the
preconditioner, and the accuracies were measured as the absolute difference
between the estimated log determinant and the log determinant as computed
using a Cholesky decomposition.  All computations were performed on Intel CPUs
with 64-bit floats and the following parameter values:
\begin{itemize}
	\item $s=35$ Rademacher probe vectors,
	\item $t=20$ iterations of the Lanczos algorithm in step 1 of Algorithm~1, and
	\item A randomized, truncated SVD preconditioner using $25$ approximate eigenvalues computed using $5$ iterations.
\end{itemize}
These parameters were selected to provide a reasonable speed/accuracy trade-off
on two problem instances representative of our intended applications:
the determinant of the covariance of Matérn-$5/2$ Gaussian process kernel
in five dimensions with $n= 20,000$, and the derivative of that determinant.
Graphs showing the sensitivity of each algorithm to each parameter
(including the choice of probe vector type and preconditioner algorithm)
can be found in the Appendix.

We also timed the algorithms on the NVidia A100 GPU \cite{9361255};
these timings are shown
in figures~6 through 9.  The $r5$ plots were again almost identical to those
of $r3$ and so were ommitted for clarity.  (The error plots are also ommited
because they are the same as when computed on the CPU.)
It should also be noted that CUDA
implementation of the Cholesky algorithm is currently faster (by a factor of
over 100) on the older V100 and P100 GPUs than on the A100, despite the A100
being much faster in general.

\begin{figure}
	\includegraphics[width=6cm]{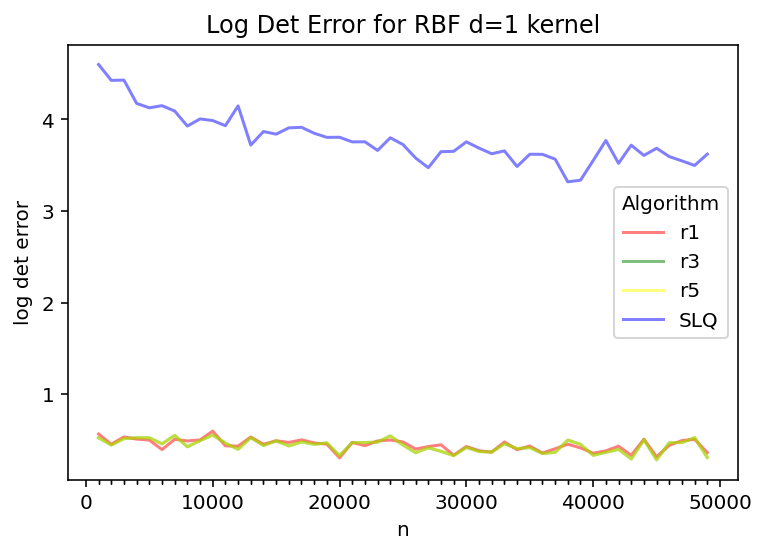}
	\includegraphics[width=6cm]{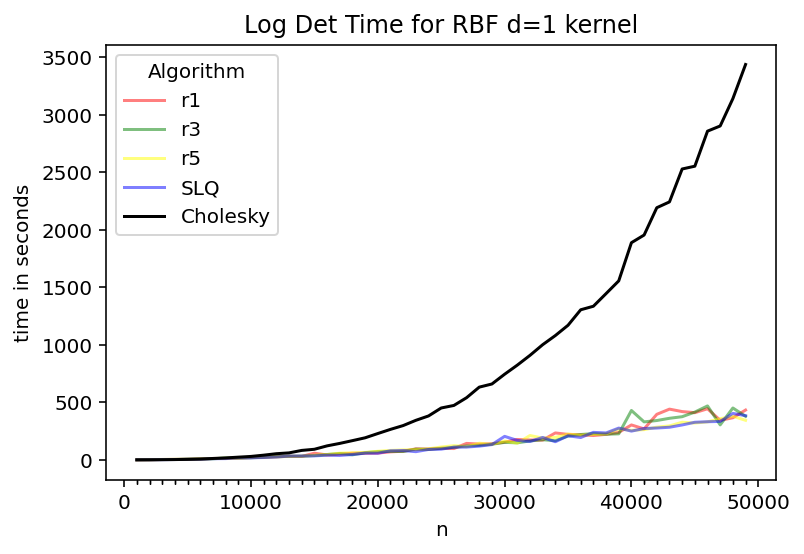}
	\caption{Comparison of $\log \det$ algorithms as a function of $n$ on the radial basis function kernel with $d=1$ as measured on an Intel CPU.  All
	measurements are averages over 100 randomly generated kernels.}
	\label{rbf1_cpu}
\end{figure}

\begin{figure}
	\includegraphics[width=6cm]{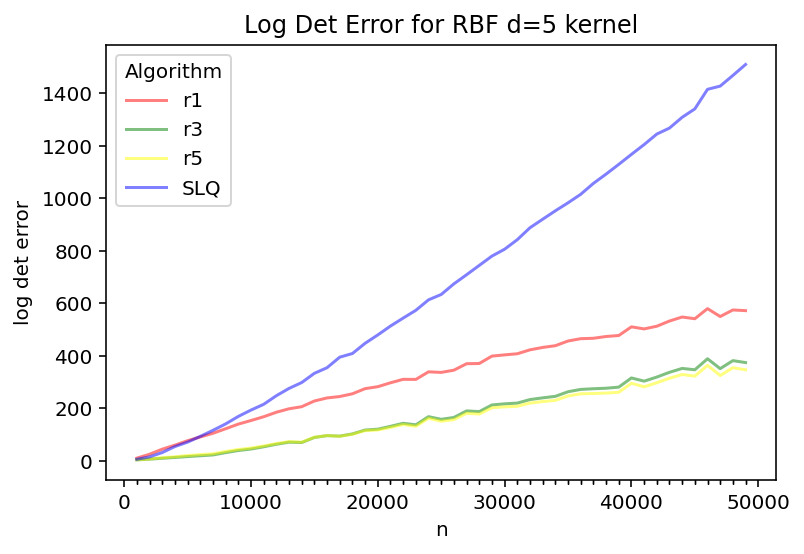}
	\includegraphics[width=6cm]{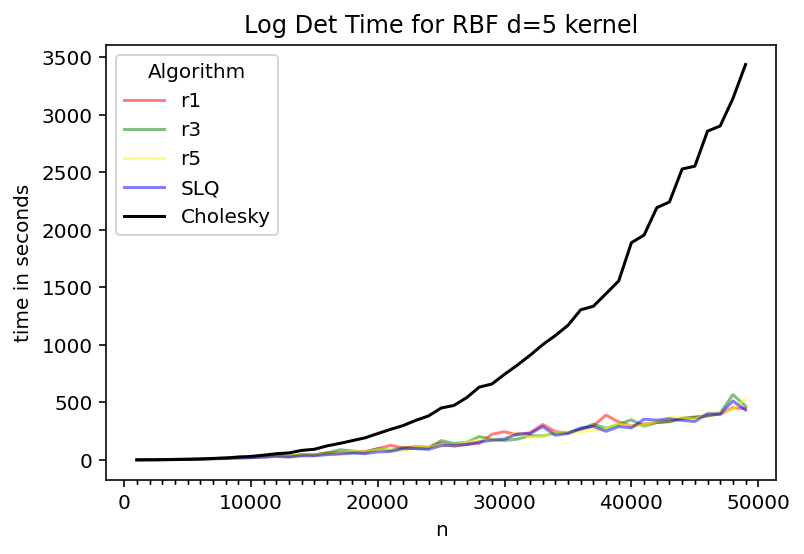}
	\caption{Comparison of $\log \det$ algorithms as a function of $n$ on the radial basis function kernel with $d=5$ as measured on an Intel CPU.  All
	measurements are averages over 100 randomly generated kernels.}
	\label{rbf5_cpu}
\end{figure}

\begin{figure}
	\includegraphics[width=6cm]{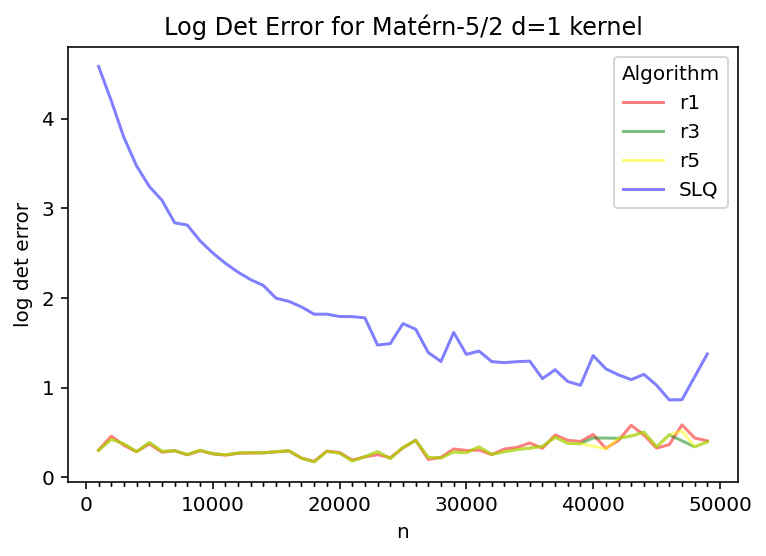}
	\includegraphics[width=6cm]{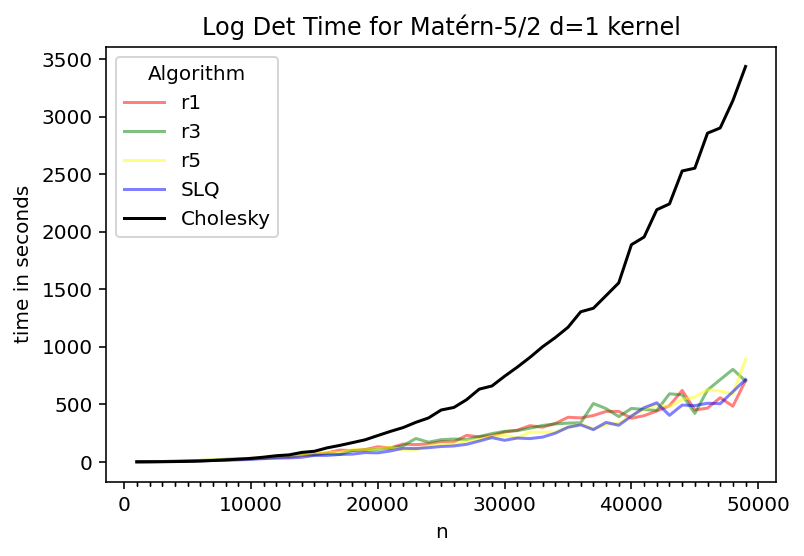}
	\caption{Comparison of $\log \det$ algorithms as a function of $n$ on the Matérn-$5/2$ kernel with $d=1$ as measured on an Intel CPU.  All
	measurements are averages over 100 randomly generated kernels.}
	\label{maternd1_cpu}
\end{figure}

\begin{figure}
	\includegraphics[width=6cm]{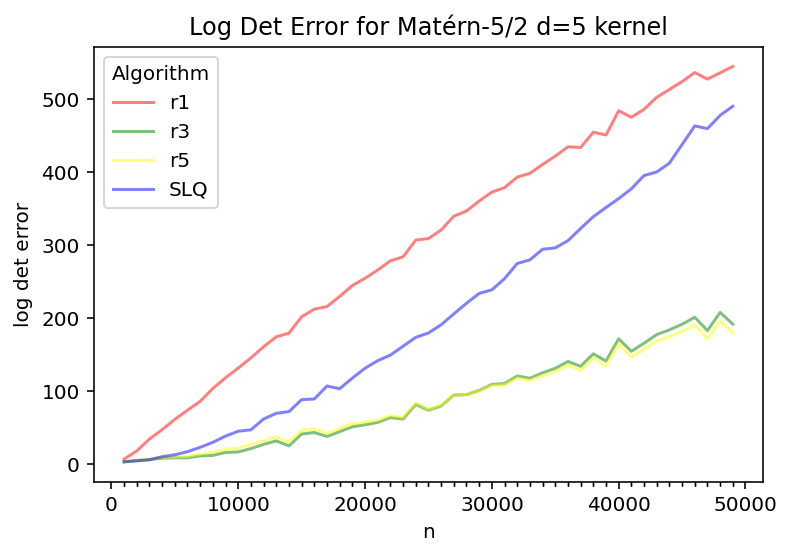}
	\includegraphics[width=6cm]{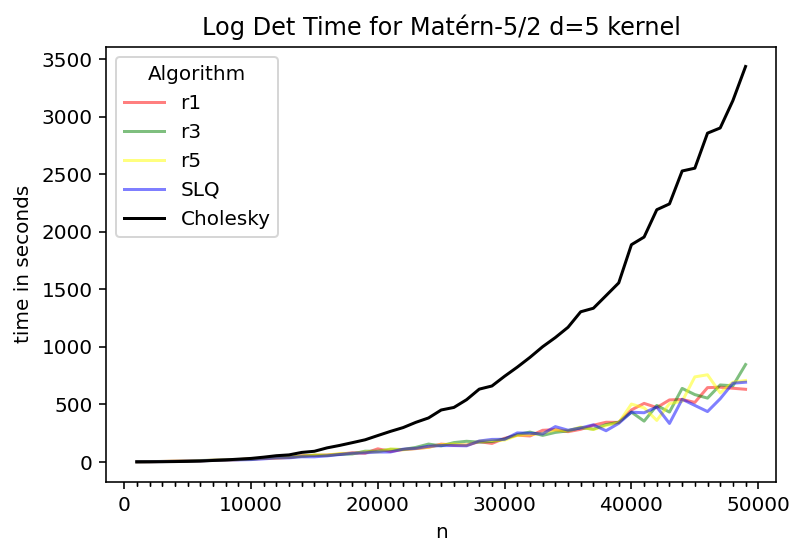}
	\caption{Comparison of $\log \det$ algorithms as a function of $n$ on the Matérn-$5/2$ kernel with $d=5$ as measured on an Intel CPU.  All
	measurements are averages over 100 randomly generated kernels.}
	\label{maternd5_cpu}
\end{figure}


\begin{figure}
	\begin{center}
	\includegraphics[width=6cm]{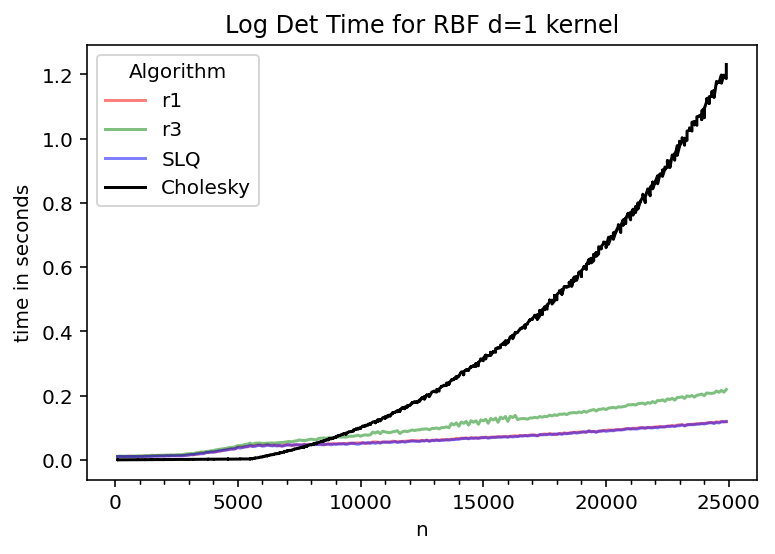}
	\end{center}
	\caption{Comparison of $\log \det$ algorithms as a function of $n$ on the radial basis function kernel with $d=1$ as measured on a NVidia A100 GPU.  All
	measurements are averages over 100 randomly generated kernels.}
	\label{rbf1}
\end{figure}

\begin{figure}
	\begin{center}
	\includegraphics[width=6cm]{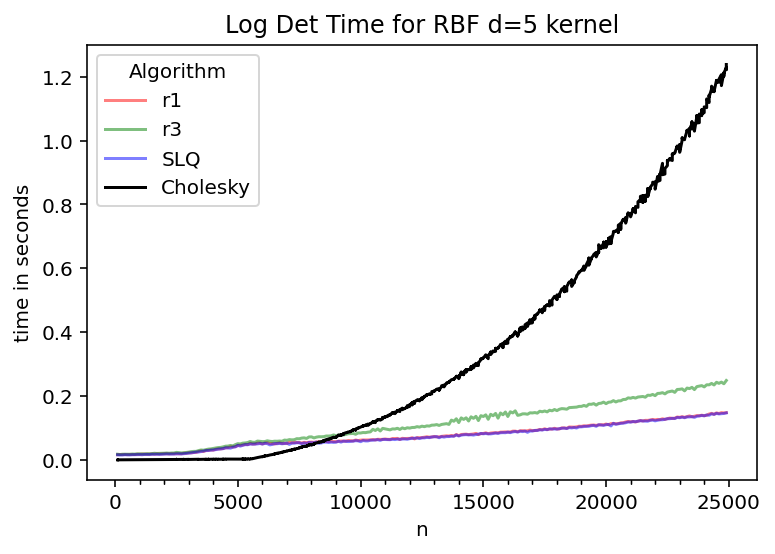}
	\end{center}
	\caption{Comparison of $\log \det$ algorithms as a function of $n$ on the radial basis function kernel with $d=5$ as measured on a NVidia A100 GPU.  All
	measurements are averages over 100 randomly generated kernels.}
	\label{rbf5}
\end{figure}

\begin{figure}
	\begin{center}
	\includegraphics[width=6cm]{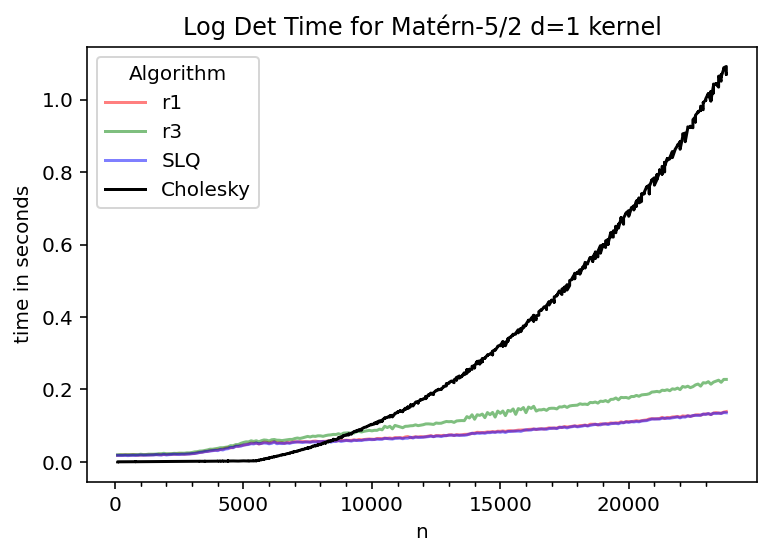}
	\end{center}
	\caption{Comparison of $\log \det$ algorithms as a function of $n$ on the Matérn-$5/2$ kernel with $d=1$ as measured on a NVidia A100 GPU.  All
	measurements are averages over 100 randomly generated kernels.}
	\label{maternd1}
\end{figure}

\begin{figure}
	\begin{center}
	\includegraphics[width=6cm]{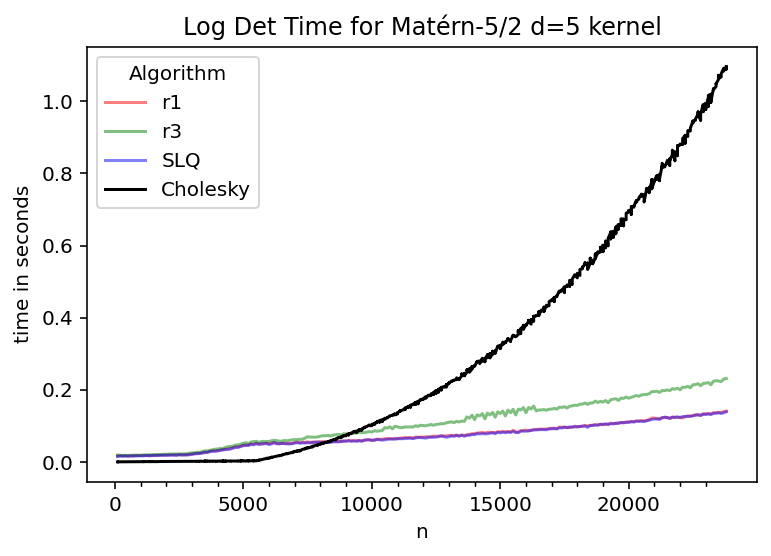}
	\end{center}
	\caption{Comparison of $\log \det$ algorithms as a function of $n$ on the Matérn-$5/2$ kernel with $d=5$ as measured on a NVidia A100 GPU.  All
	measurements are averages over 100 randomly generated kernels.}
	\label{maternd5}
\end{figure}

From these plots, we make the following observations:
\begin{itemize}
	\item The r3 and r5 algorithms consistently have the lowest errors
		over the four kernel types and matrix sizes (up to 50,000)
		investigated.  It is particularly noteworthy that r5 does
		not have a noticeably lower error than r3, despite being a
		closer approximation to $\log M$ in theory.
	\item All of the r* and SLQ algorithms have approximately the same
		running time when measured on Intel CPUs.  When measured on
		NVidia A100 GPUs, the r1 and SLQ algorithms have almost
		exactly the same running time and are slightly faster than
		the r3 algorithm.
	\item The underlying dimension "d" of the Gaussian process has an
		extremely large impact on the accuracy of the $\log \det$
		approximation algorithms.  For 50,000 by 50,000 matrices
		for example, the absolute errors of the r3 and r5 algorithms
		on the $d=5$ kernels are over 400 times that of their errors
		on the $d=1$ kernels.
\end{itemize}

\begin{figure}
	\begin{center}
	\includegraphics[width=6cm]{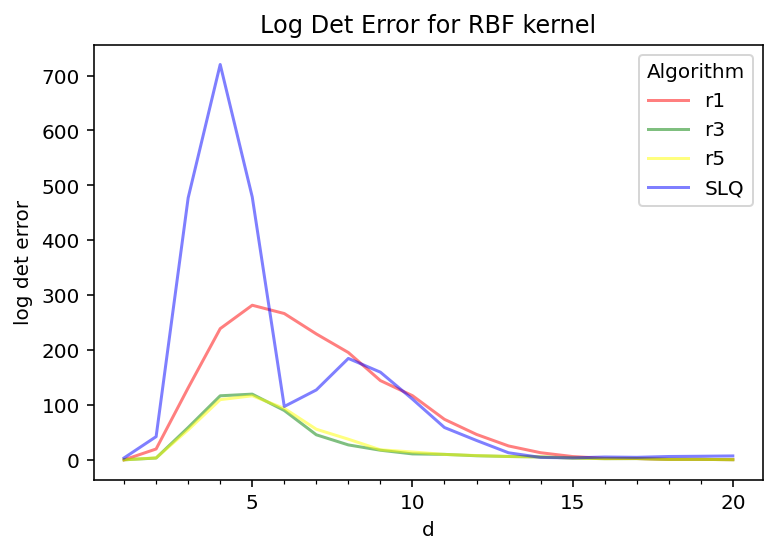}
	\includegraphics[width=6cm]{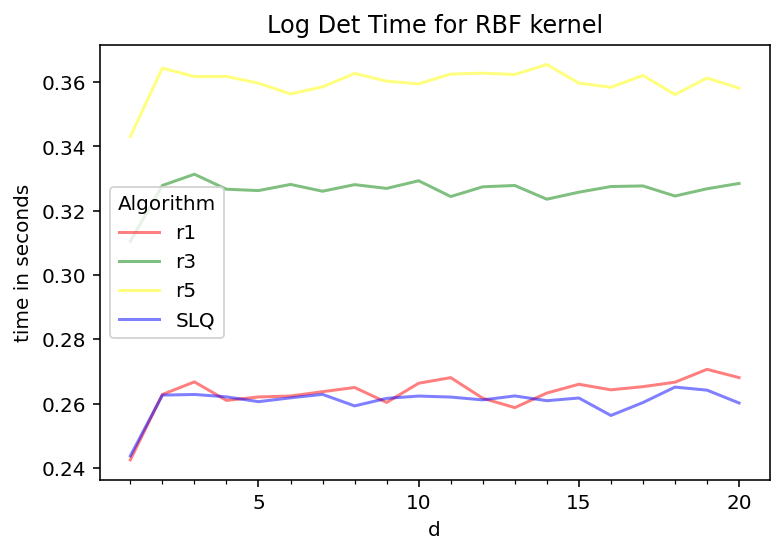}
	\includegraphics[width=6cm]{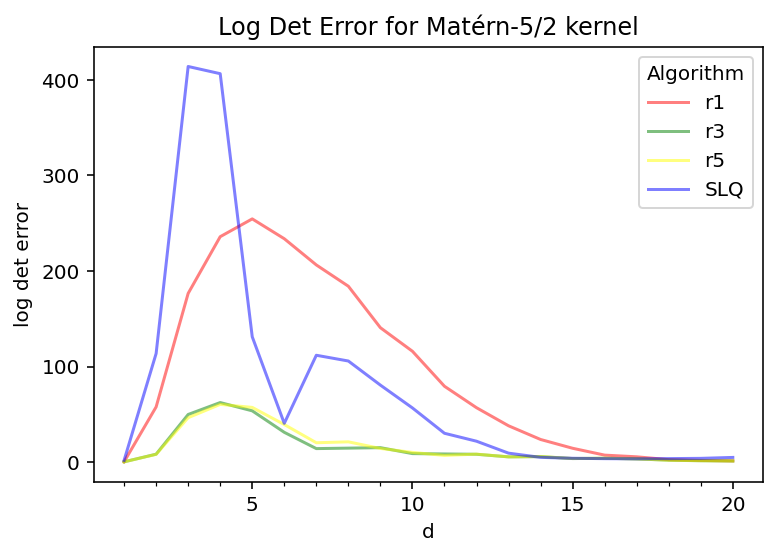}
	\includegraphics[width=6cm]{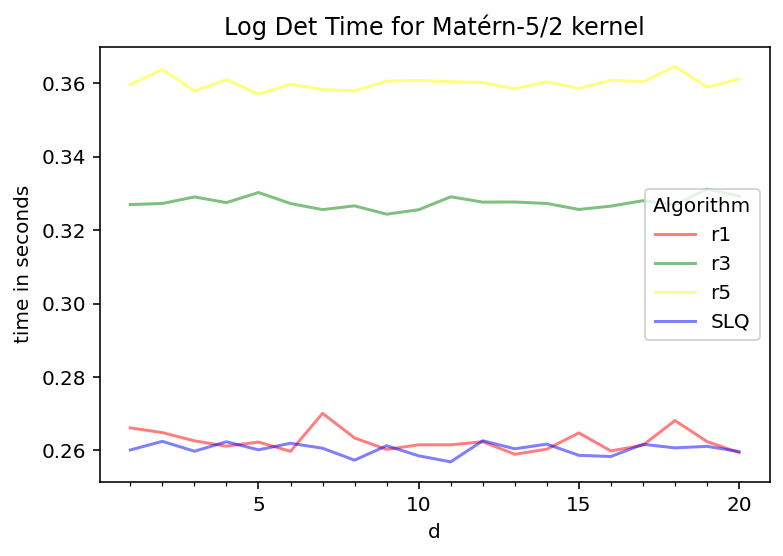}
	\end{center}
	\caption{Comparison of $\log \det$ algorithms when run on Gaussian proceess covariance matrices with different underlying dimension "d".  All measurements are averages over 100 randomly generated kernels with $n = 20,000$ as measured on a NVidia A100 GPU.}
	\label{d_sweep}
\end{figure}

To understand that last item more deeply, we ran a sweep over different
"d" values while holding the matrix size fixed at $n = 20,000$; the results
are presented in Figure~\ref{d_sweep}.  For both RBF and Matérn-$5/2$ kernels,
absolute errors were highest for intermediate values of $d$ centered around
$d=5$ and lowest for $d=1$ and $d > 15$.  The more accurate r3 and r5
algorithms had shorter and narrower error peaks than the r1 and SLQ
algorithms, with SLQ having the highest error peaks overall.

\section{Conclusions}

We have presented an algorithm for computing a new family of approximations
to the matrix determinant.  This algorithm combines classical rational
function approximations to $\log x$ with well known techniques like
Hutchinson's trace estimator and the novel (in the space of determinant
approximation algorithms) union of partial fraction decompositions and
fast multi-shift solvers.
In our results, one member of this family, r3,
consistently achieved a lower error than the state of the art stochastic
lanczos quadrature approximation, with only a slightly higher running time.
The accuracy advantage of r3 over SLQ was particularly significant when
measured on covariance matrices coming from Gaussian process kernels with
underlying dimension greater than one.

It would be interesting for future work to examine whether these patterns
hold over a wider class of matrix families.  We are also curious as to why for
all of the examined algorithms it appears harder to approximate the
determinant of a covariance matrix derived from points in the moderate
underlying dimension "d" range of $3$ to $15$ than it is for the $d < 3$
or $d > 15$ cases (Figure~\ref{d_sweep}).  Based on
Figure~\ref{preconditioner_rank_sweep}, it appears that the preconditioner
behaves oddly in those dimensions, with the error first increasing as the
preconditioner size increases, and then decreasing more slowly than the
theoretical work of \cite{wenger2022preconditioning} would suggest.


\section{Bibliography}

\bibliographystyle{plain}
\bibliography{references}

\begin{thebibliography}{10}

\bibitem{boutsidis2017randomized}
Christos Boutsidis, Petros Drineas, Prabhanjan Kambadur, Eugenia-Maria
  Kontopoulou, and Anastasios Zouzias.
\newblock A randomized algorithm for approximating the log determinant of a
  symmetric positive definite matrix.
\newblock {\em Linear Algebra and its Applications}, 533:95--117, 2017.

\bibitem{jax2018github}
James Bradbury, Roy Frostig, Peter Hawkins, Matthew~James Johnson, Chris Leary,
  Dougal Maclaurin, George Necula, Adam Paszke, Jake Vander{P}las, Skye
  Wanderman-{M}ilne, and Qiao Zhang.
\newblock {JAX}: composable transformations of {P}ython+{N}um{P}y programs,
  2018.

\bibitem{9361255}
Jack Choquette, Wishwesh Gandhi, Olivier Giroux, Nick Stam, and Ronny
  Krashinsky.
\newblock Nvidia a100 tensor core gpu: Performance and innovation.
\newblock {\em IEEE Micro}, 41(2):29--35, 2021.

\bibitem{dillon2017tensorflow}
Joshua~V Dillon, Ian Langmore, Dustin Tran, Eugene Brevdo, Srinivas Vasudevan,
  Dave Moore, Brian Patton, Alex Alemi, Matt Hoffman, and Rif~A Saurous.
\newblock Tensorflow distributions.
\newblock {\em arXiv preprint arXiv:1711.10604}, 2017.

\bibitem{ebden2015gaussian}
Mark Ebden.
\newblock Gaussian processes: A quick introduction.
\newblock {\em arXiv preprint arXiv:1505.02965}, 2015.

\bibitem{epperly2023}
Ethan~N. Epperly.
\newblock Stochastic trace estimation.
\newblock \url
  {https://www.ethanepperly.com/index.php/2023/01/26/stochastic-trace-estimation/},
  2023.
\newblock Accessed 2024-04-17.

\bibitem{gimeno2017computation}
Joan Gimeno~Alqu{\'e}zar.
\newblock On computation of matrix logarithm times a vector.
\newblock Master's thesis, Universitat Polit{\`e}cnica de Catalunya, 2017.

\bibitem{halko2011finding}
Nathan Halko, Per-Gunnar Martinsson, and Joel~A Tropp.
\newblock Finding structure with randomness: Probabilistic algorithms for
  constructing approximate matrix decompositions.
\newblock {\em SIAM review}, 53(2):217--288, 2011.

\bibitem{han2015large}
Insu Han, Dmitry Malioutov, and Jinwoo Shin.
\newblock Large-scale log-determinant computation through stochastic chebyshev
  expansions.
\newblock In {\em International Conference on Machine Learning}, pages
  908--917. PMLR, 2015.

\bibitem{hutchinson}
M.F. Hutchinson.
\newblock A stochastic estimator of the trace of the influence matrix for
  laplacian smoothing splines.
\newblock {\em Communications in Statistics - Simulation and Computation},
  19(2):433--450, 1990.

\bibitem{jegerlehner1996krylov}
Beat Jegerlehner.
\newblock Krylov space solvers for shifted linear systems.
\newblock {\em arXiv preprint hep-lat/9612014}, 1996.

\bibitem{kelisky1968rational}
RP~Kelisky and TJ~Rivlin.
\newblock A rational approximation to the logarithm.
\newblock {\em Mathematics of Computation}, 22(101):128--136, 1968.

\bibitem{knyazev2001toward}
Andrew~V Knyazev.
\newblock Toward the optimal preconditioned eigensolver: Locally optimal block
  preconditioned conjugate gradient method.
\newblock {\em SIAM journal on scientific computing}, 23(2):517--541, 2001.

\bibitem{lanczos1950iteration}
Cornelius Lanczos.
\newblock An iteration method for the solution of the eigenvalue problem of
  linear differential and integral operators.
\newblock 1950.

\bibitem{lee2003zone}
Dean~J Lee and Ilse~CF Ipsen.
\newblock Zone determinant expansions for nuclear lattice simulations.
\newblock {\em Physical review C}, 68(6):064003, 2003.

\bibitem{martinsson2020randomized}
Per-Gunnar Martinsson and Joel~A Tropp.
\newblock Randomized numerical linear algebra: Foundations and algorithms.
\newblock {\em Acta Numerica}, 29:403--572, 2020.

\bibitem{PACE2004179}
R.~Kelley Pace and James~P. LeSage.
\newblock Chebyshev approximation of log-determinants of spatial weight
  matrices.
\newblock {\em Computational Statistics \& Data Analysis}, 45(2):179--196,
  2004.

\bibitem{Strassen1973}
Volker Strassen.
\newblock Vermeidung von divisionen.
\newblock {\em Journal für die reine und angewandte Mathematik}, 264:184--202,
  1973.

\bibitem{strassen1969gaussian}
Volker Strassen et~al.
\newblock Gaussian elimination is not optimal.
\newblock {\em Numerische mathematik}, 13(4):354--356, 1969.

\bibitem{thomas1949elliptic}
Llewellyn~Hilleth Thomas.
\newblock Elliptic problems in linear difference equations over a network.
\newblock {\em Watson Sci. Comput. Lab. Rept., Columbia University, New York},
  1:71, 1949.

\bibitem{ubaru2017fast}
Shashanka Ubaru, Jie Chen, and Yousef Saad.
\newblock Fast estimation of tr(f(a)) via stochastic lanczos quadrature.
\newblock {\em SIAM Journal on Matrix Analysis and Applications},
  38(4):1075--1099, 2017.

\bibitem{wang2019exact}
Ke~Wang, Geoff Pleiss, Jacob Gardner, Stephen Tyree, Kilian~Q Weinberger, and
  Andrew~Gordon Wilson.
\newblock Exact gaussian processes on a million data points.
\newblock {\em Advances in neural information processing systems}, 32, 2019.

\bibitem{wenger2022preconditioning}
Jonathan Wenger, Geoff Pleiss, Philipp Hennig, John Cunningham, and Jacob
  Gardner.
\newblock Preconditioning for scalable gaussian process hyperparameter
  optimization.
\newblock In {\em International Conference on Machine Learning}, pages
  23751--23780. PMLR, 2022.

\bibitem{williams2006gaussian}
Christopher~KI Williams and Carl~Edward Rasmussen.
\newblock {\em Gaussian processes for machine learning}, volume~2.
\newblock MIT press Cambridge, MA, 2006.

\bibitem{yu2016orthogonal}
Felix Xinnan~X Yu, Ananda~Theertha Suresh, Krzysztof~M Choromanski, Daniel~N
  Holtmann-Rice, and Sanjiv Kumar.
\newblock Orthogonal random features.
\newblock {\em Advances in neural information processing systems}, 29, 2016.

\bibitem{zhang2007approximate}
Yunong Zhang and William~E Leithead.
\newblock Approximate implementation of the logarithm of the matrix determinant
  in gaussian process regression.
\newblock {\em Journal of Statistical Computation and Simulation},
  77(4):329--348, 2007.

\end{thebibliography}

\section{Appendix}

In this section, we investigate the sensitivity of our $\log \det$
approximation algorithms to their hyperparameters.

\begin{figure}
	\begin{center}
	\includegraphics[width=6cm]{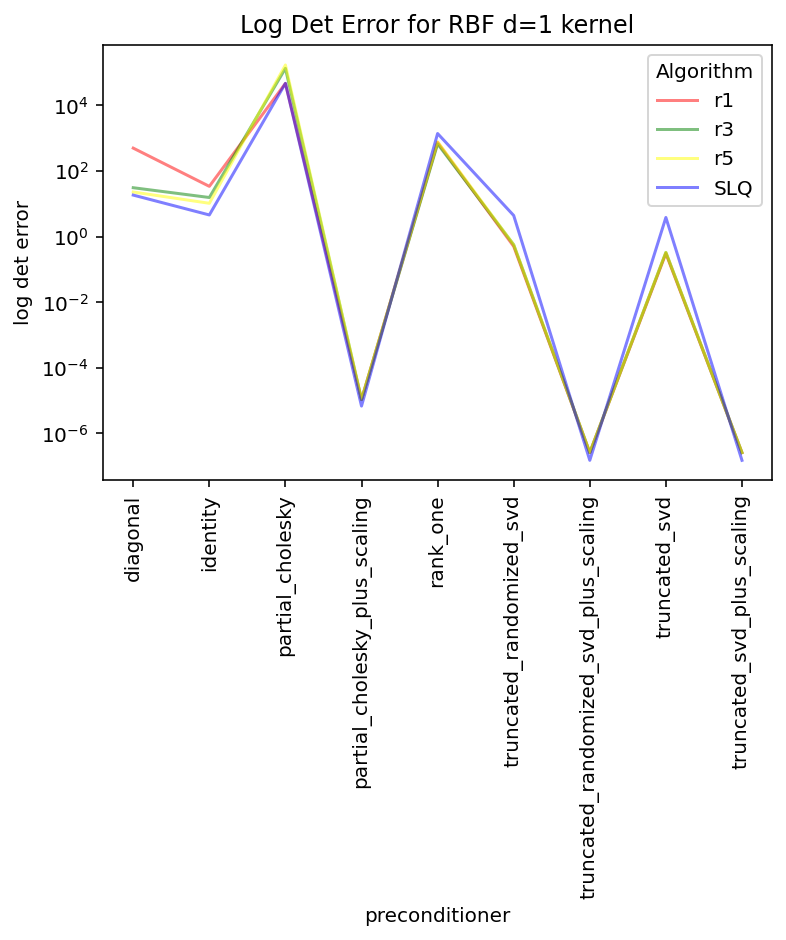}
	\includegraphics[width=6cm]{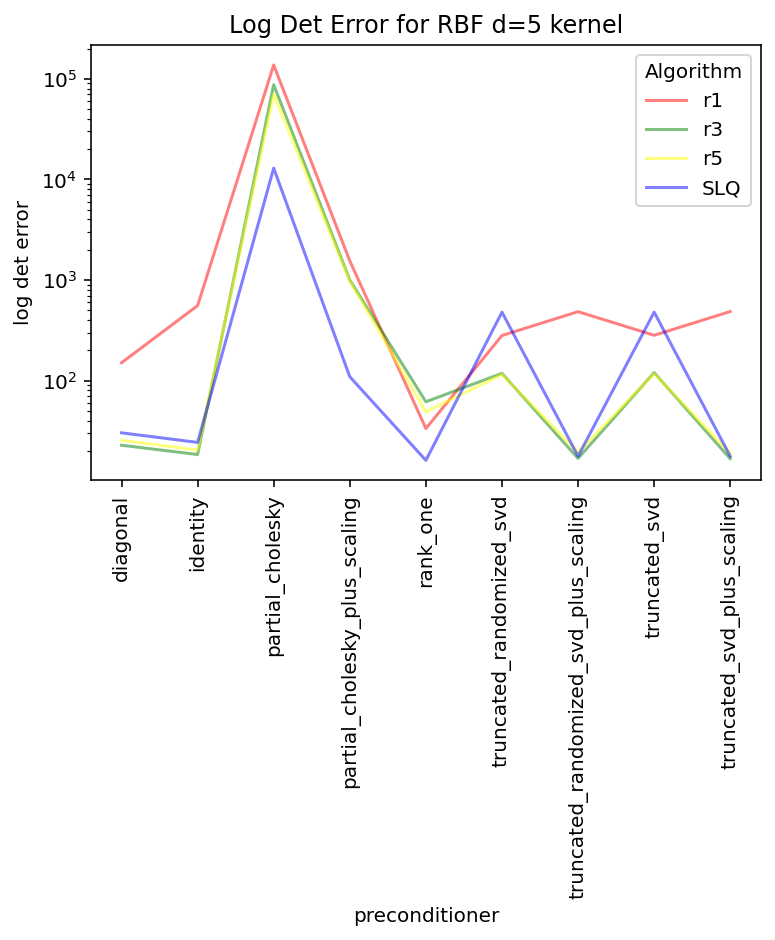}
	\includegraphics[width=6cm]{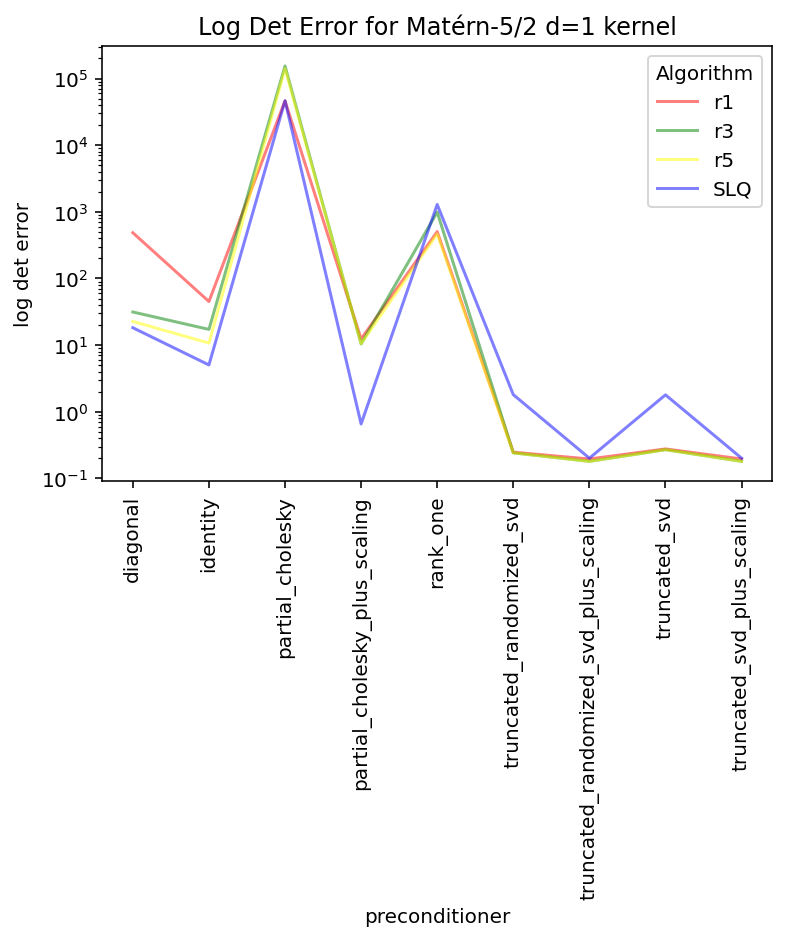}
	\includegraphics[width=6cm]{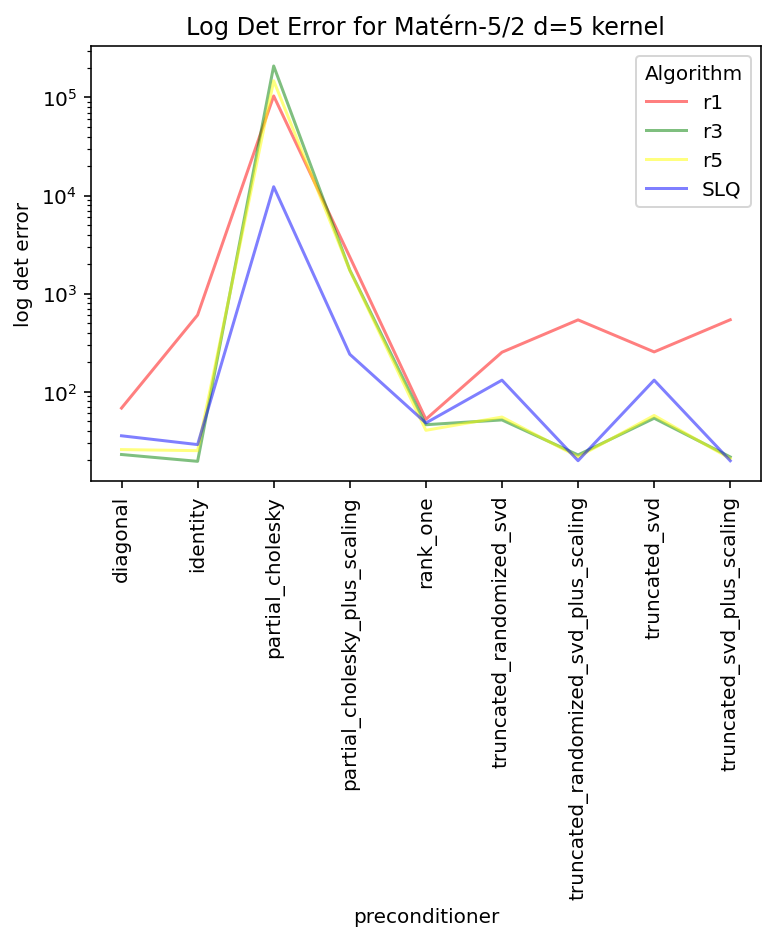}
	\end{center}
	\caption{Comparison of $\log \det$ algorithm accuracies when run with different preconditioners.  All measurements are averages over 100 randomly generated kernels with $n = 20,000$ as measured on a NVidia A100 GPU.}
	\label{preconditioner_sweep_errors}
\end{figure}

\begin{figure}
	\begin{center}
	\includegraphics[width=6cm]{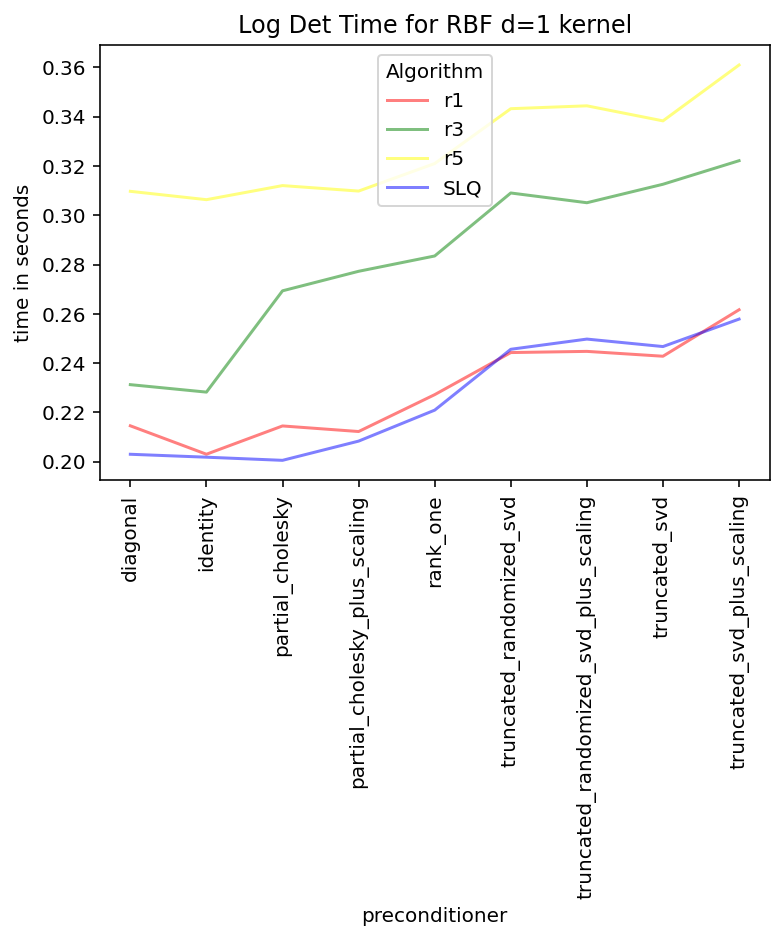}
	\includegraphics[width=6cm]{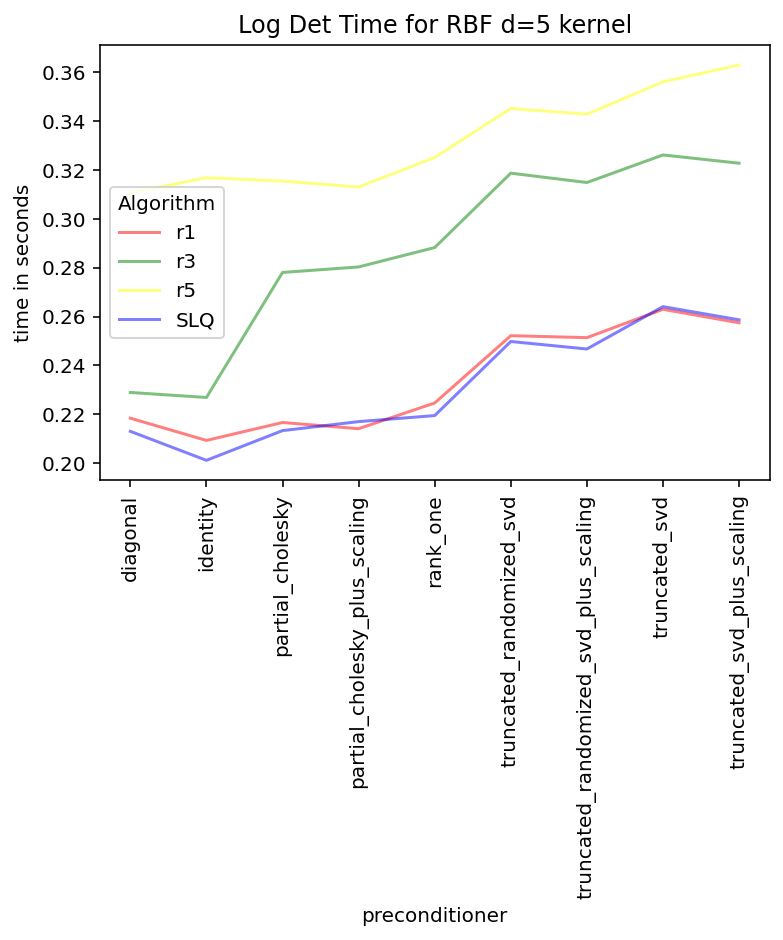}
	\includegraphics[width=6cm]{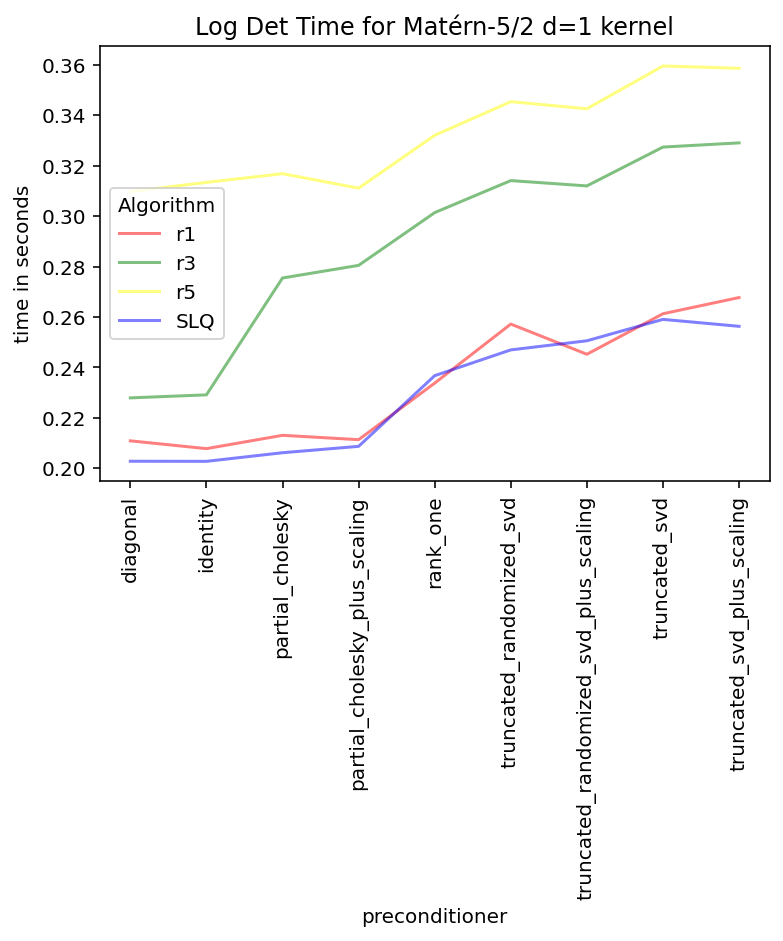}
	\includegraphics[width=6cm]{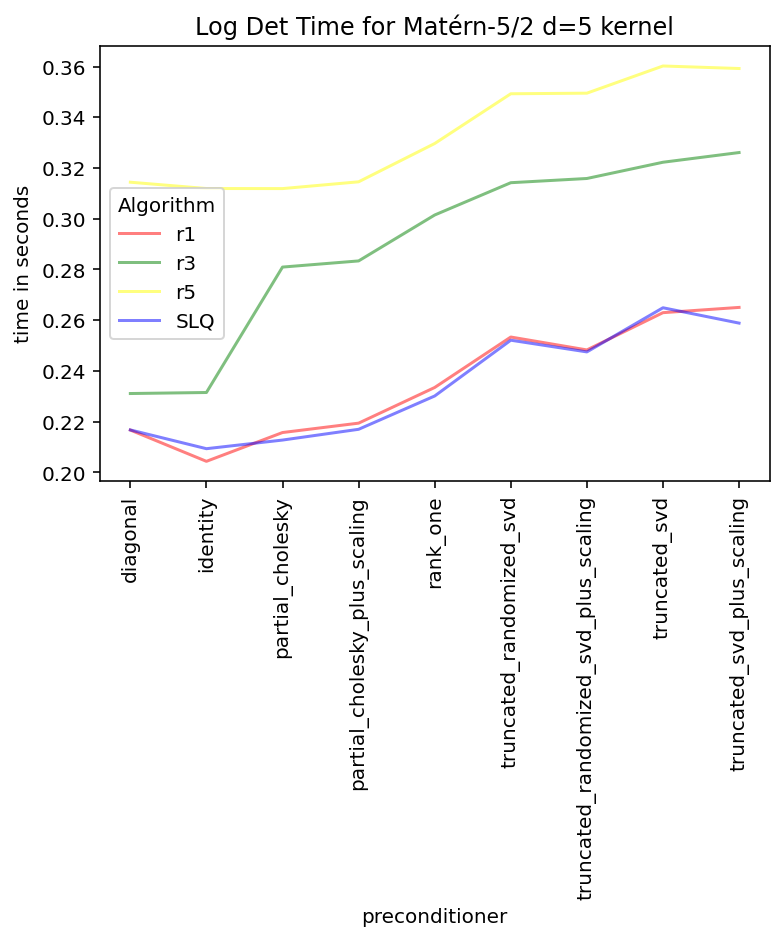}
	\end{center}
	\caption{Comparison of $\log \det$ algorithm running times when run with different preconditioners.  All measurements are averages over 100 randomly generated kernels with $n = 20,000$ as measured on a NVidia A100 GPU.}
	\label{preconditioner_sweep_times}
\end{figure}

\begin{figure}
	\begin{center}
	\includegraphics[width=6cm]{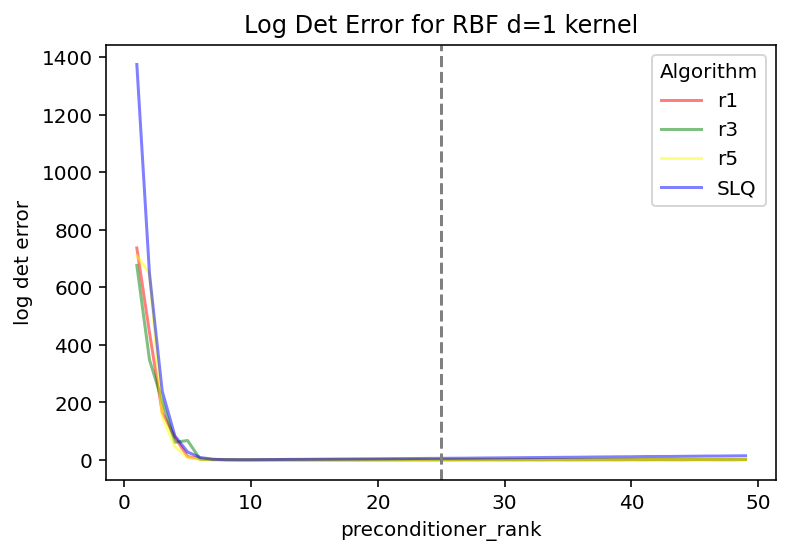}
	\includegraphics[width=6cm]{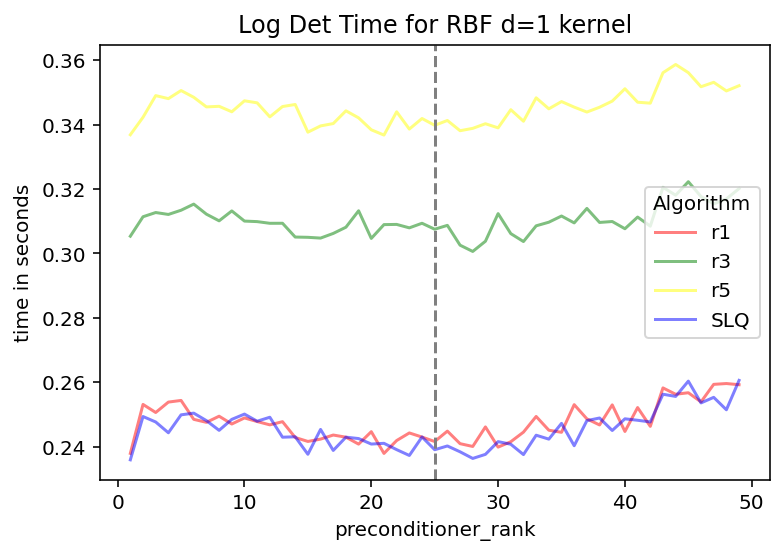}
	\includegraphics[width=6cm]{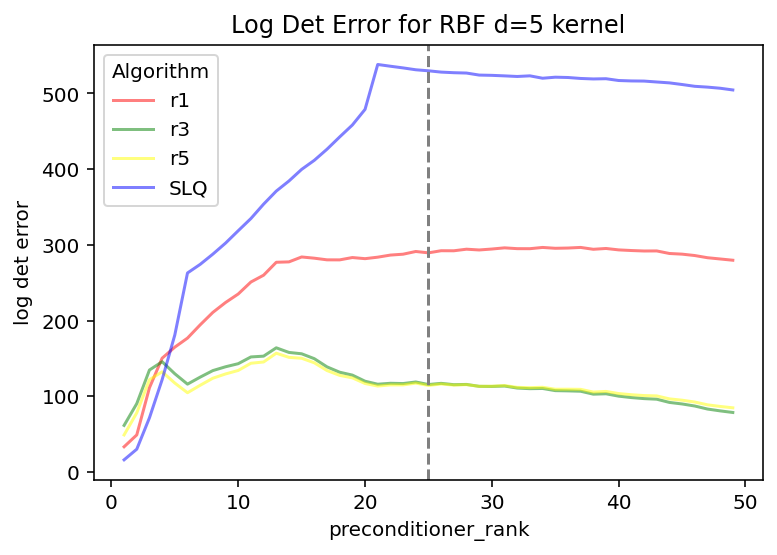}
	\includegraphics[width=6cm]{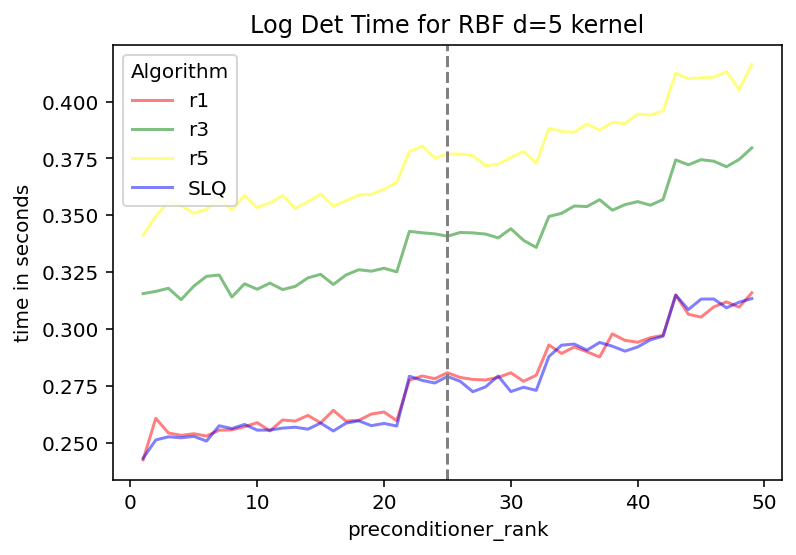}
	\includegraphics[width=6cm]{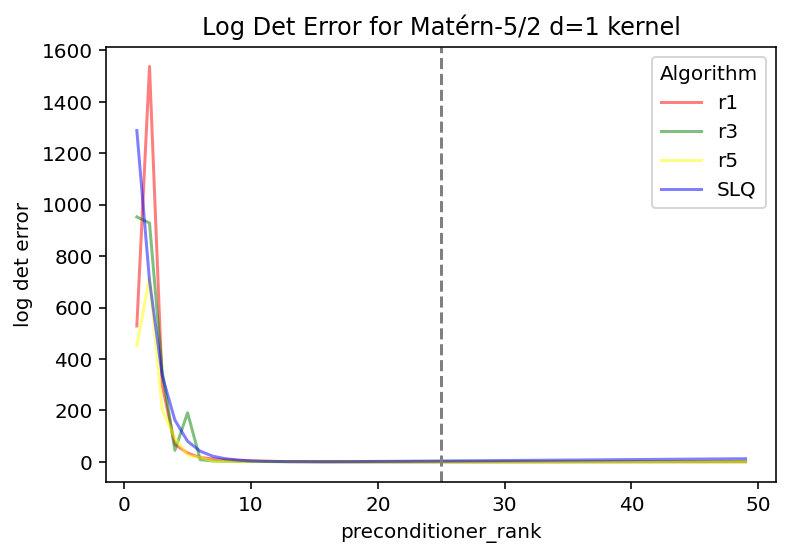}
	\includegraphics[width=6cm]{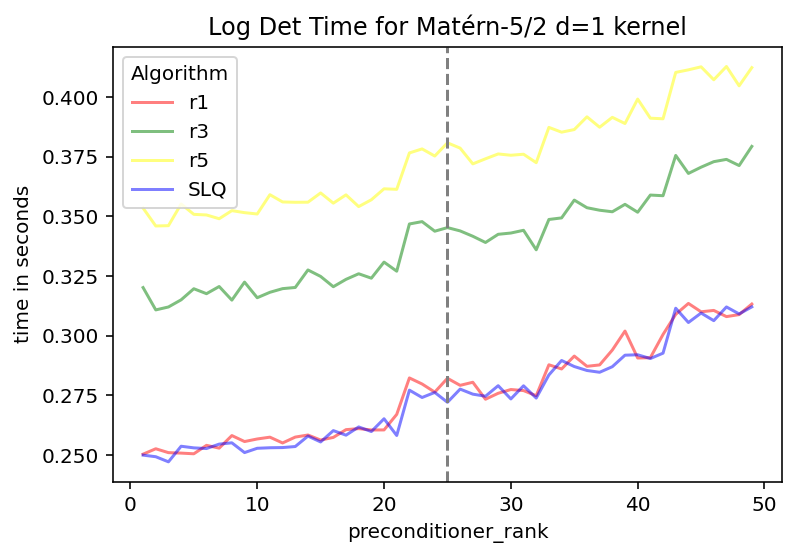}
	\includegraphics[width=6cm]{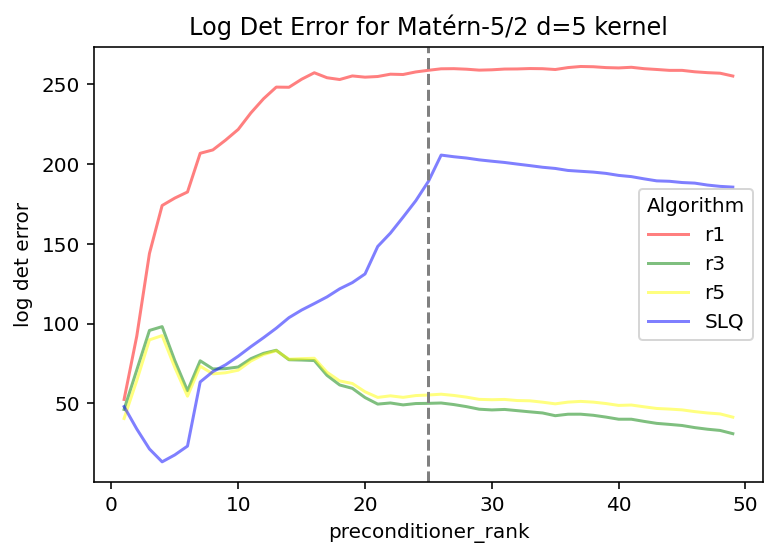}
	\includegraphics[width=6cm]{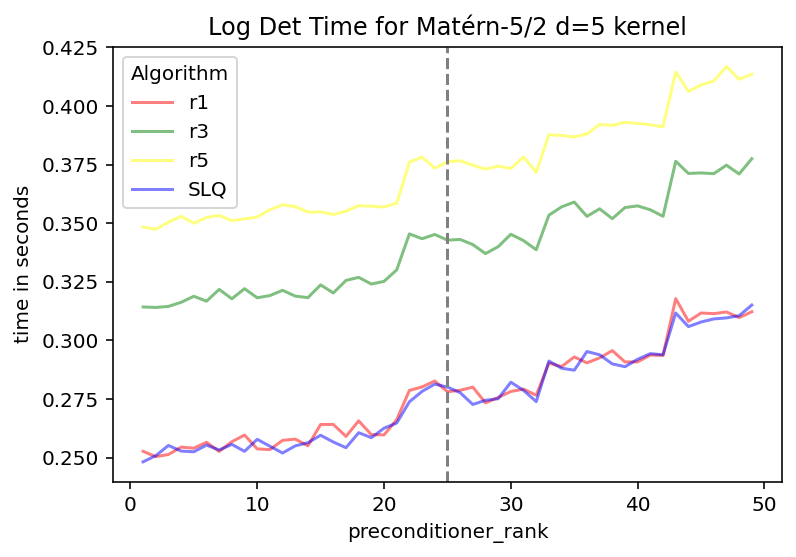}
	\end{center}
	\caption{Comparison of $\log \det$ algorithms when run with differently sized preconditioners.  All measurements are averages over 100 randomly generated kernels with $n = 20,000$ as measured on a NVidia A100 GPU.}
	\label{preconditioner_rank_sweep}
\end{figure}

Figures~\ref{preconditioner_sweep_errors} and \ref{preconditioner_sweep_times}
show their performance as a function of the preconditioner used.  The
preconditioners examined are
\begin{itemize}
	\item {\bf Identity}: $P(M) = I$.
	\item {\bf Diagonal}: $P(M) = \mbox{diag}(M)$.
	\item {\bf Rank One}: $P(M) = D + \lambda v v^t$ where $\lambda$ is
		$M$'s largest eigenvalue, $v$ is the corresponding eigenvector,
		and $D$ is a diagonal matrix
		$D = \mbox{diag}(M - \lambda v v^t)$.
	\item {\bf Partial Cholesky}: $P(M) = D + C C^t$ where $C C^t$ is the
		incomplete Cholesky factorization of $M$ of rank
		${\tt preconditioner\_rank}$ and
		$D = \mbox{diag}(M - \lambda C D^t)$.
	\item {\bf Partial Cholesky plus scaling}: $P(M) = aI + C C^t$ where
		$C C^t$ is the
                incomplete Cholesky factorization of $M$ of rank
                ${\tt preconditioner\_rank}$ and
		$a$ is the sum of the Gaussian process's jitter and
		observation noise variance parameters.
	\item {\bf Truncated SVD}: $P(M) = D + A A^t$ where
		$A A^t$ is the matrix formed by $M$'s top
		${\tt preconditioner\_rank}$ standard eigenvalues as computed
		by the Locally Optimal Block Preconditioned Conjugated
		Gradient algorithm \cite{knyazev2001toward} and implemented in
		{\tt jax.experimental.sparse.linalg.lobpcg\_standard}.
	\item {\bf Truncated SVD plus scaling}: Same as above, but with
		$P(M) = aI + A A^t$.
	\item {\bf Truncated Randomized SVD}: $P(M) = D + A A^t$ where
		$A$ is the approximate SVD described in
		\cite{halko2011finding}.
	\item {\bf Truncated Randomized SVD plus scaling}: Same as above,
		but with $P(M) = aI + A A^t$.
\end{itemize}

The 'normal orthogonal' probe vector type is an application of Orthogonal Monte Carlo \cite{yu2016orthogonal}. These probe vectors are generated via the following process:
$W_ORF = \frac{1}{\sigma}SQ$, where $\sigma$ is the variance of the Normal distribution (here $\sigma = 1$), $S$ is a diagonal matrix filled with i.i.d random variables
sampled from a $\chi$-distribution with $D$ degrees of freedom, and $Q$ is a $D$ by $s$ random orthogonal matrix. As proven in \cite{yu2016orthogonal}, each column of $W_ORF$
is marginally distributed as a spherical multivariate normal. By sampling in this way, we can enforce orthogonality on the probe vectors, which can often be used to reduce variance.


\begin{figure}
	\begin{center}
	\includegraphics[width=6cm]{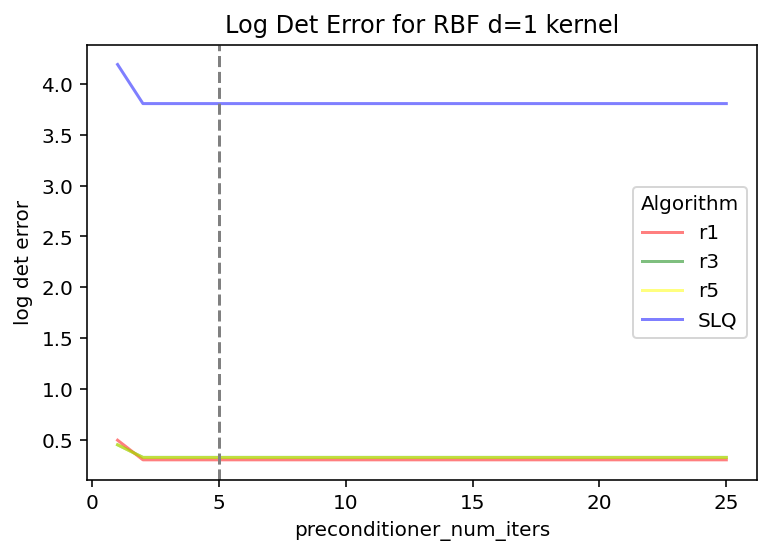}
	\includegraphics[width=6cm]{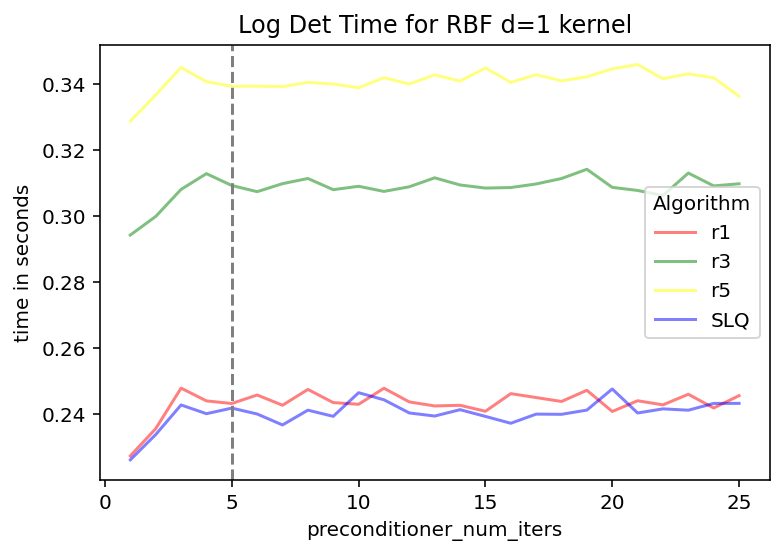}
	\includegraphics[width=6cm]{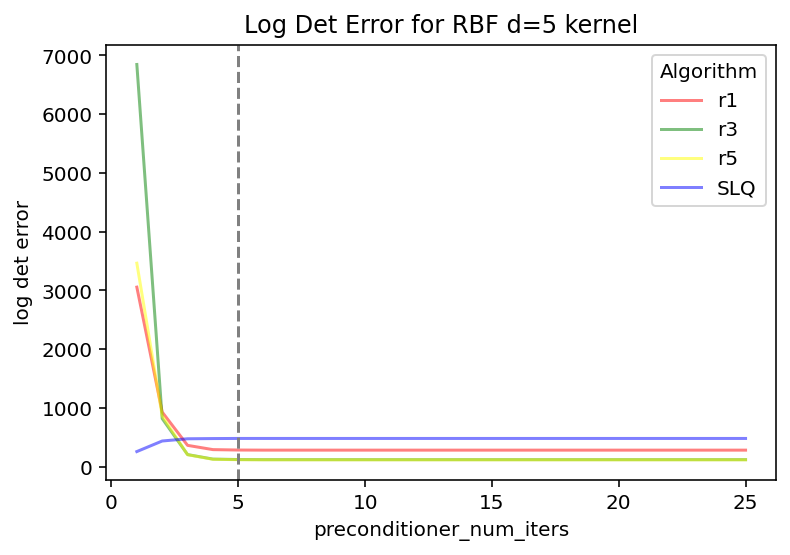}
	\includegraphics[width=6cm]{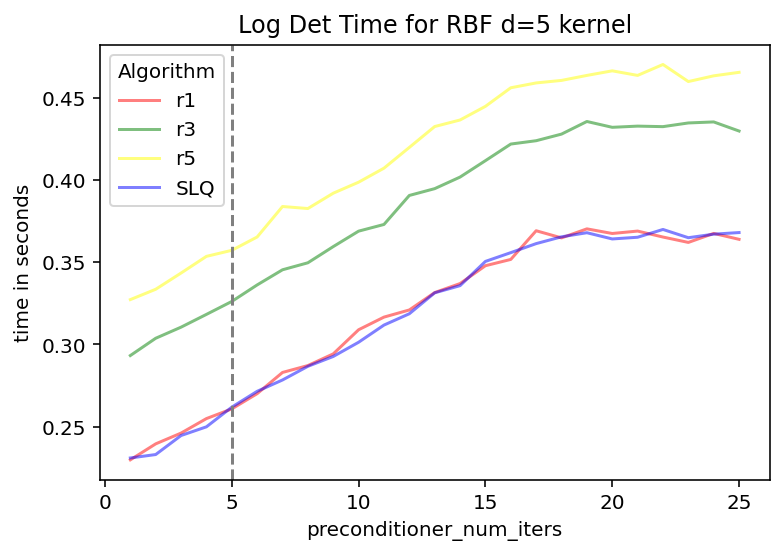}
	\includegraphics[width=6cm]{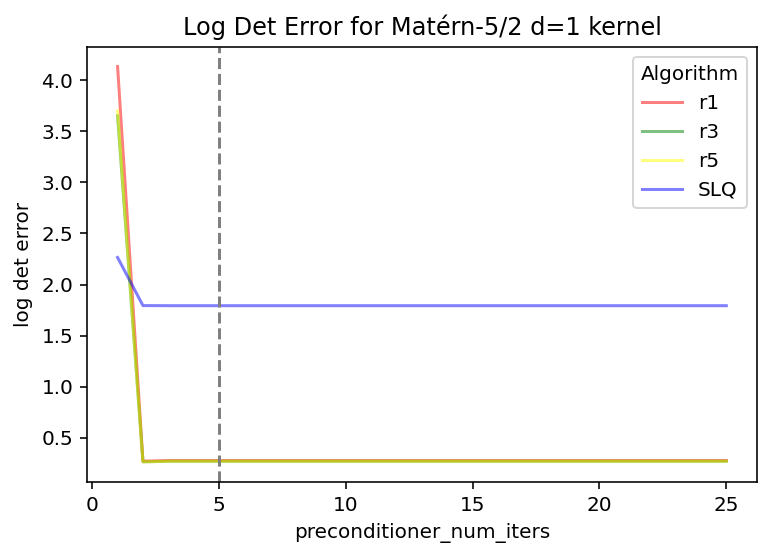}
	\includegraphics[width=6cm]{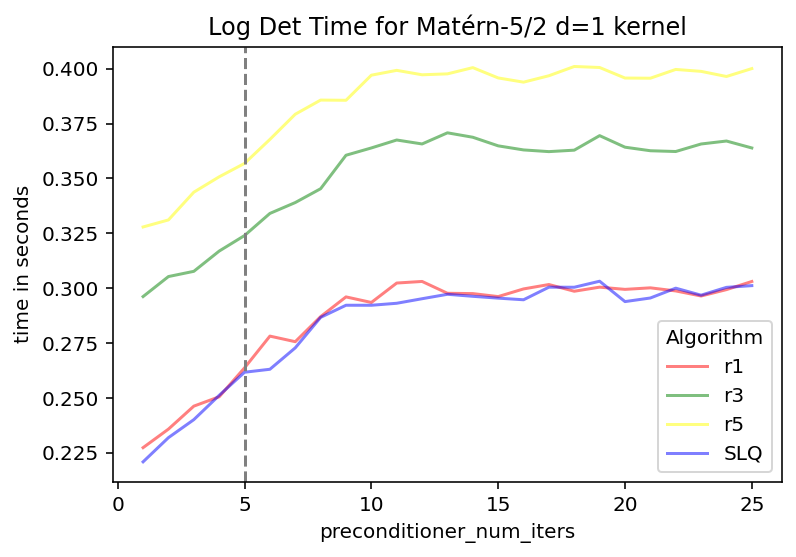}
	\includegraphics[width=6cm]{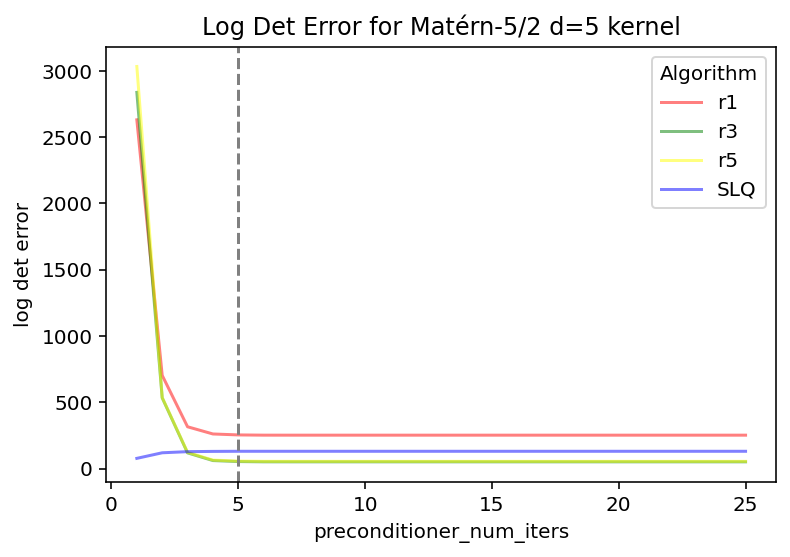}
	\includegraphics[width=6cm]{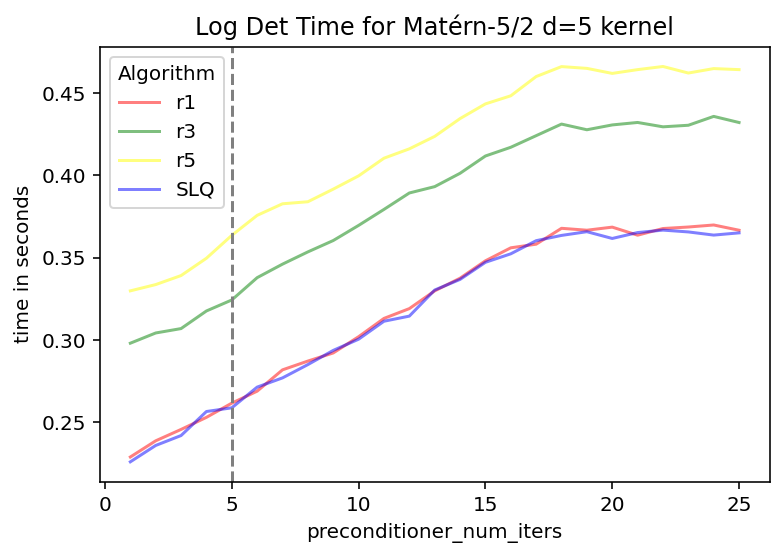}
	\end{center}
	\caption{Comparison of $\log \det$ algorithms when run with different values of the {\tt preconditioner\_num\_iters} parameter.  All measurements are averages over 100 randomly generated kernels with $n = 20,000$ as measured on a NVidia A100 GPU.}
	\label{preconditioner_num_iters_sweep}
\end{figure}

\begin{figure}
	\begin{center}
	\includegraphics[width=6cm]{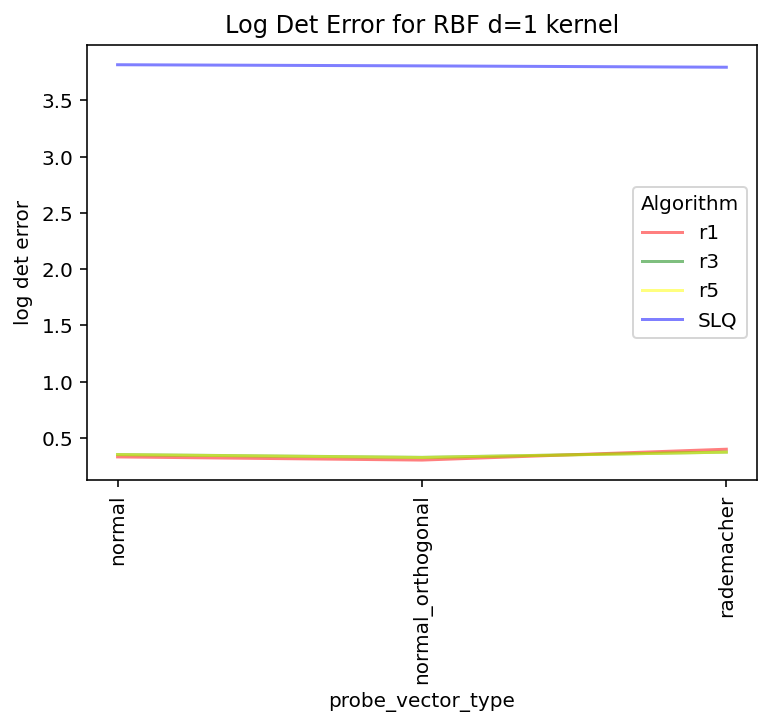}
	\includegraphics[width=6cm]{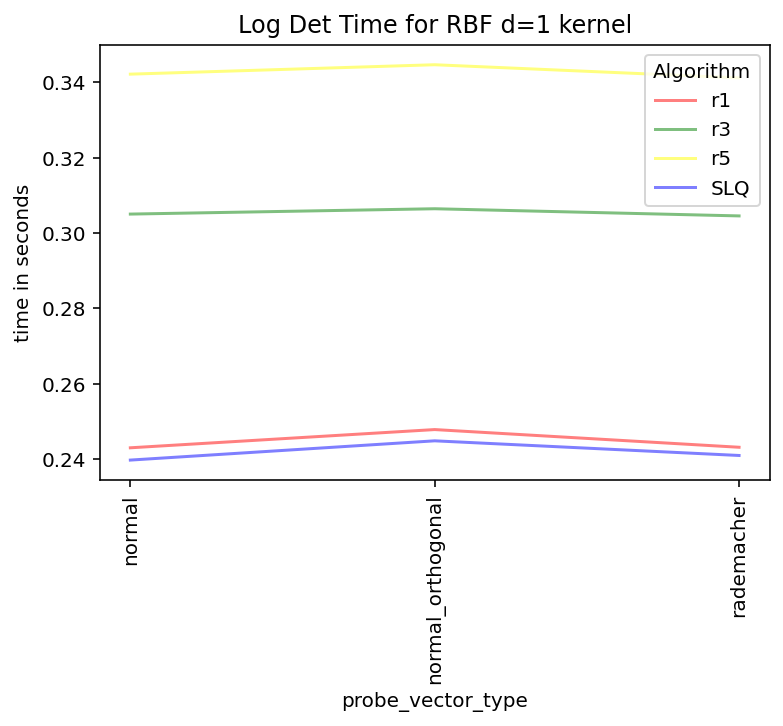}
	\includegraphics[width=6cm]{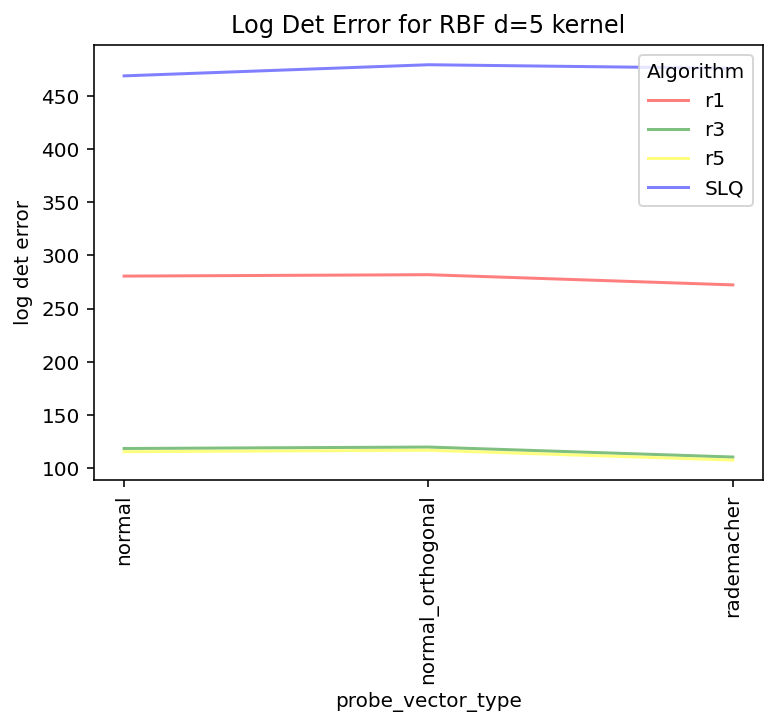}
	\includegraphics[width=6cm]{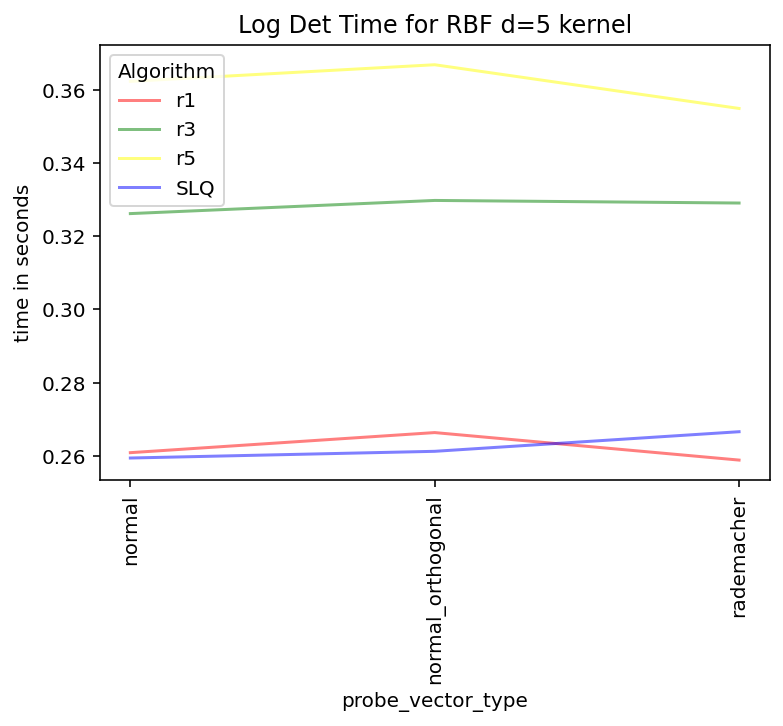}
	\includegraphics[width=6cm]{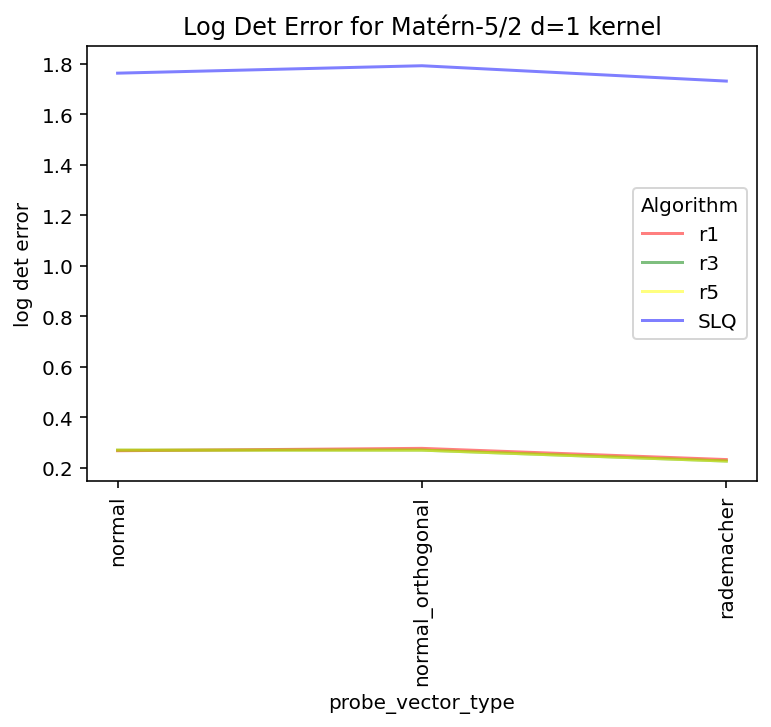}
	\includegraphics[width=6cm]{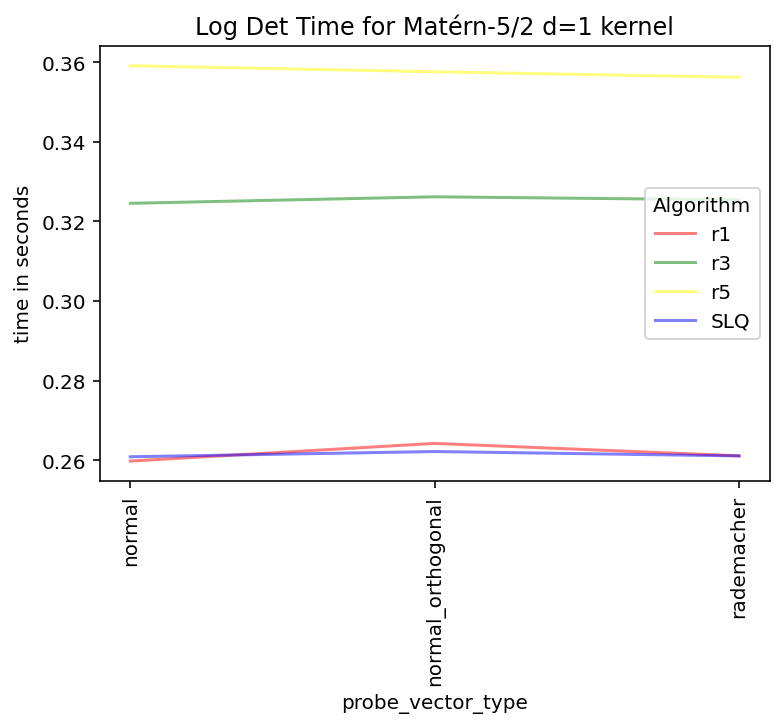}
	\includegraphics[width=6cm]{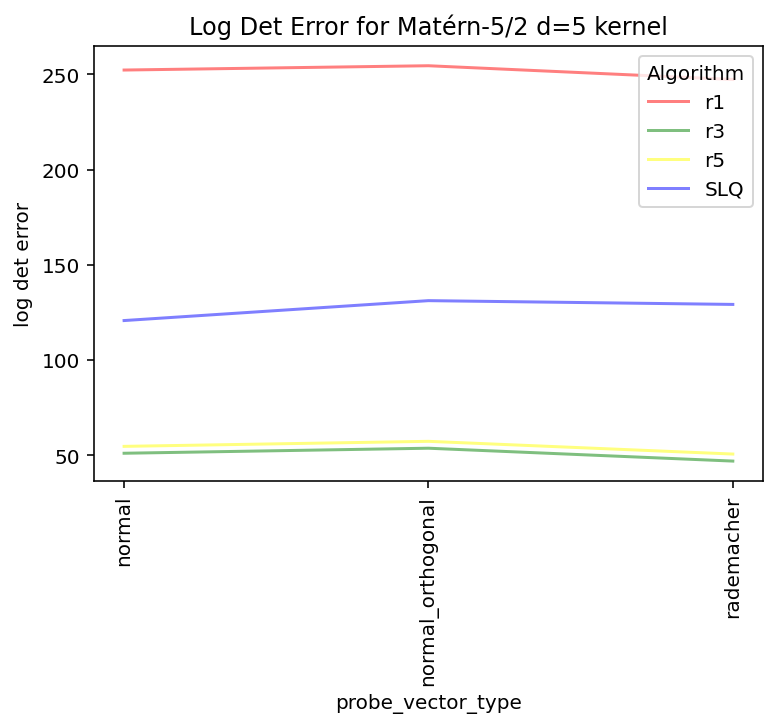}
	\includegraphics[width=6cm]{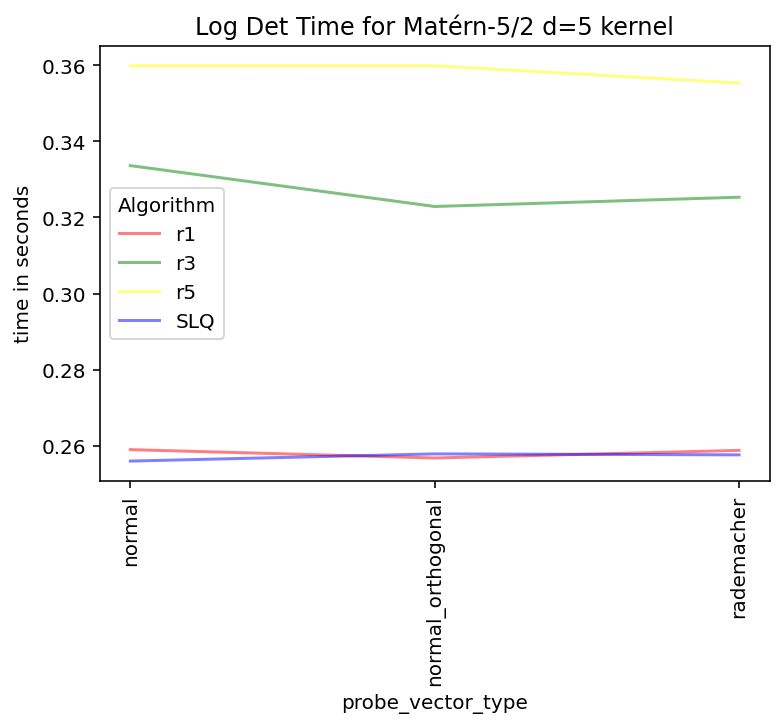}
	\end{center}
	\caption{Comparison of $\log \det$ algorithms when run with different types of probe vectors.  All measurements are averages over 100 randomly generated kernels with $n = 20,000$ as measured on a NVidia A100 GPU.}
	\label{probe_vector_type_sweep}
\end{figure}

\begin{figure}
	\begin{center}
	\includegraphics[width=6cm]{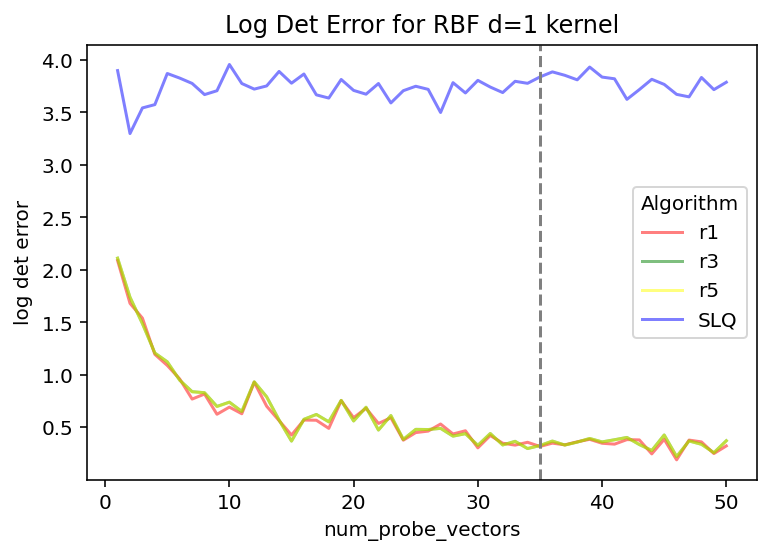}
	\includegraphics[width=6cm]{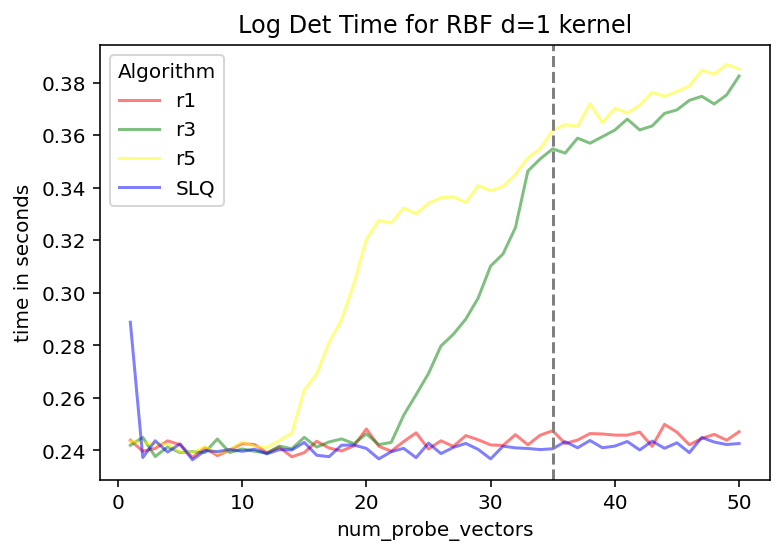}
	\includegraphics[width=6cm]{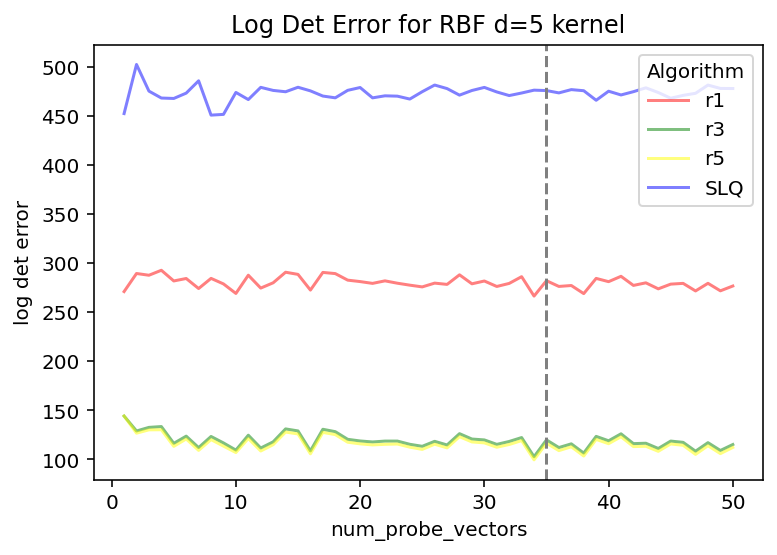}
	\includegraphics[width=6cm]{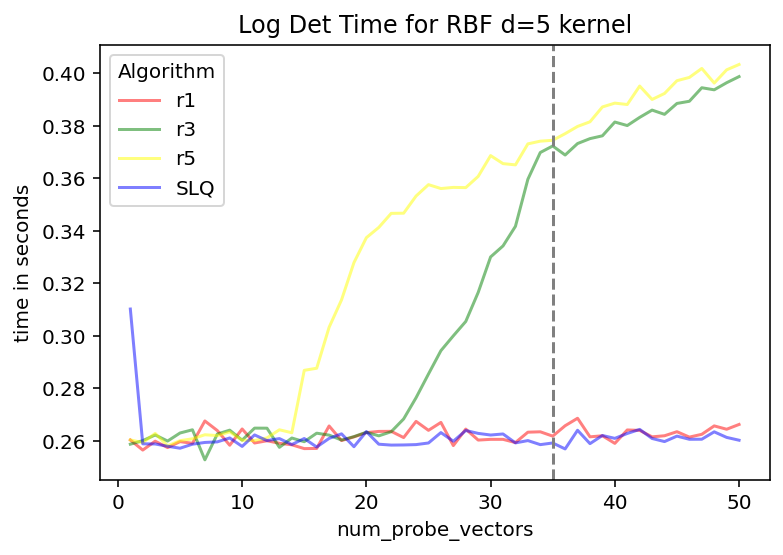}
	\includegraphics[width=6cm]{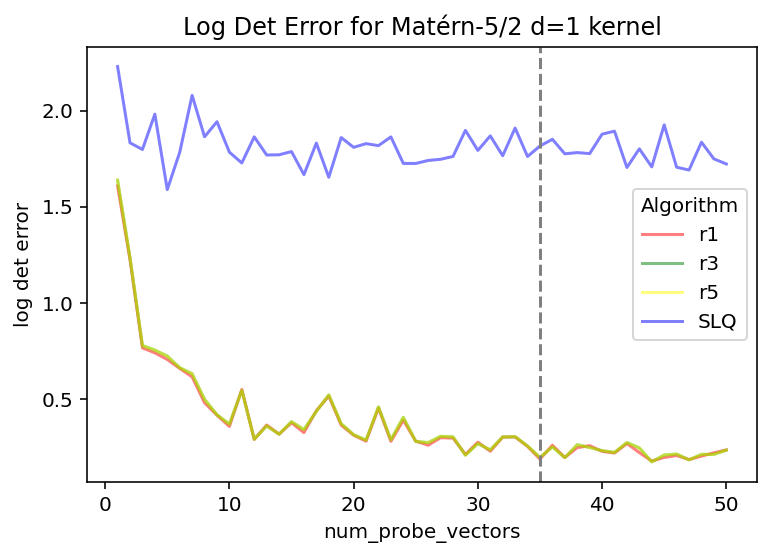}
	\includegraphics[width=6cm]{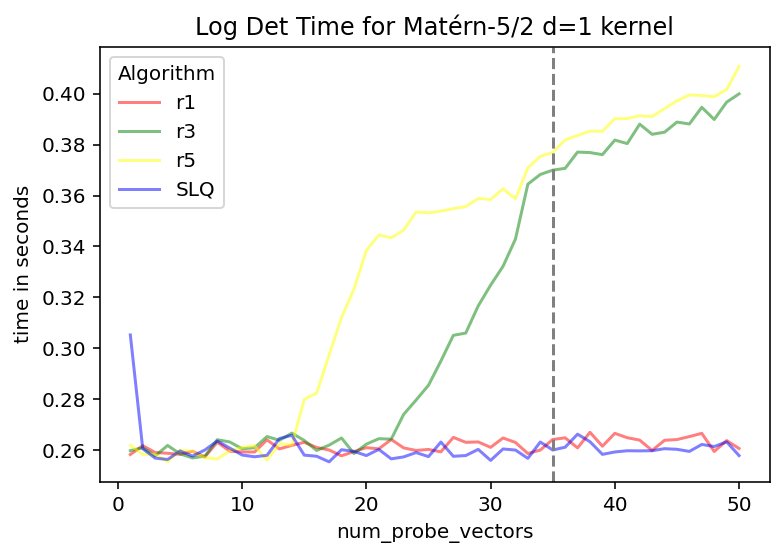}
	\includegraphics[width=6cm]{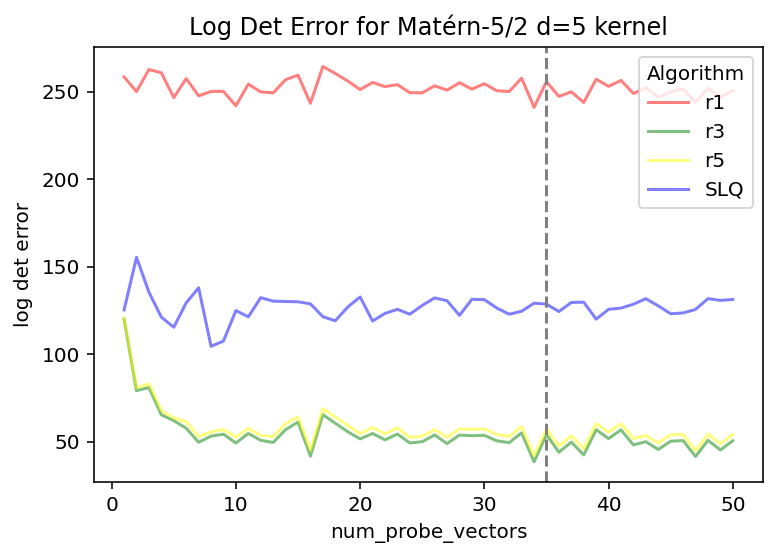}
	\includegraphics[width=6cm]{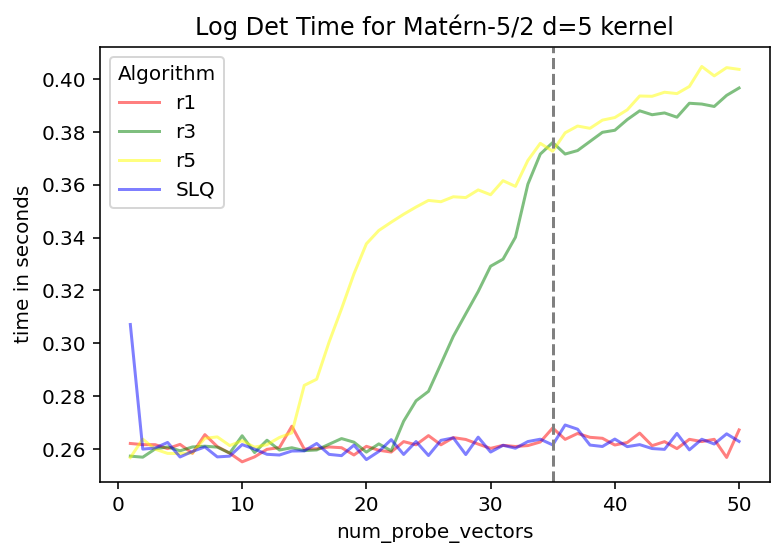}
	\end{center}
	\caption{Comparison of $\log \det$ algorithms when run with different numbers of probe vectors.  All measurements are averages over 100 randomly generated kernels with $n = 20,000$ as measured on a NVidia A100 GPU.}
	\label{num_probe_vectors_sweep}
\end{figure}

\begin{figure}
	\begin{center}
	\includegraphics[width=6cm]{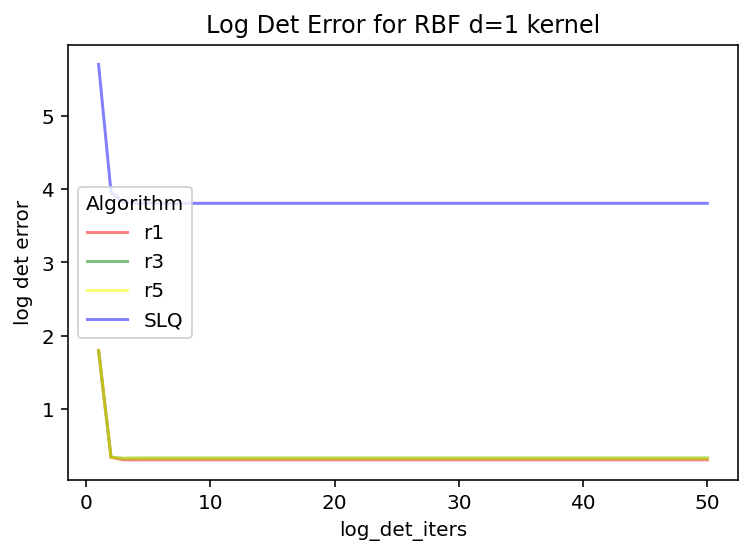}
	\includegraphics[width=6cm]{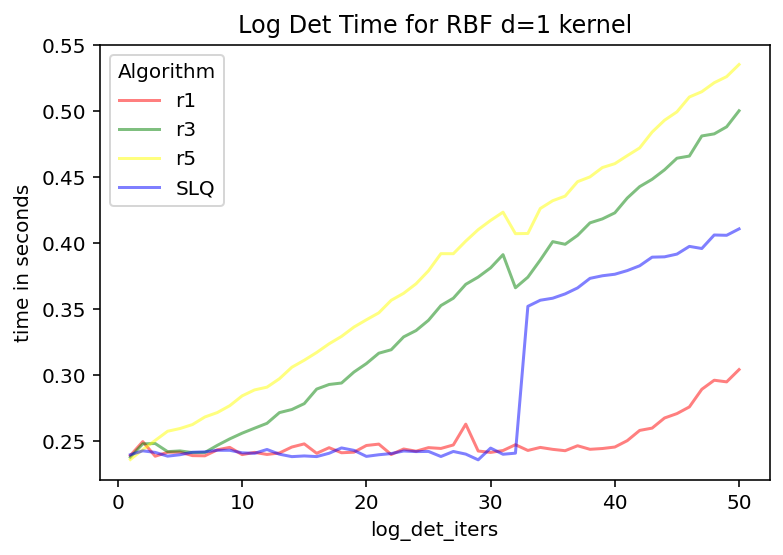}
	\includegraphics[width=6cm]{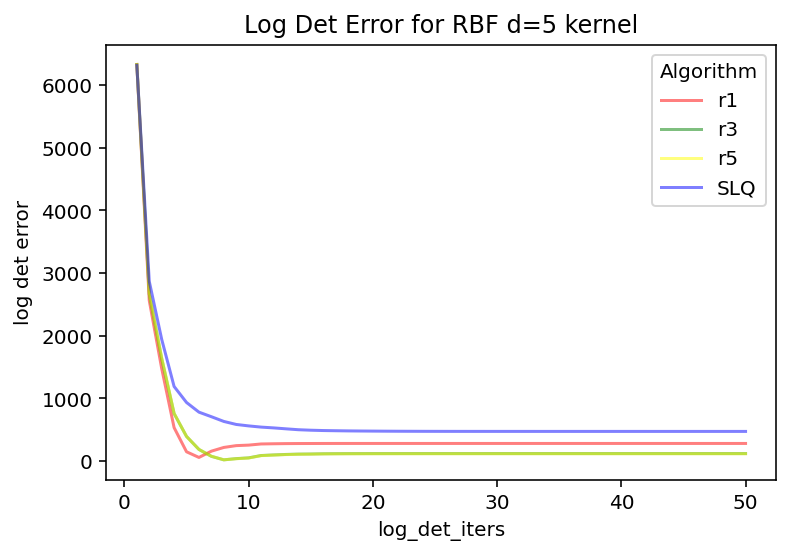}
	\includegraphics[width=6cm]{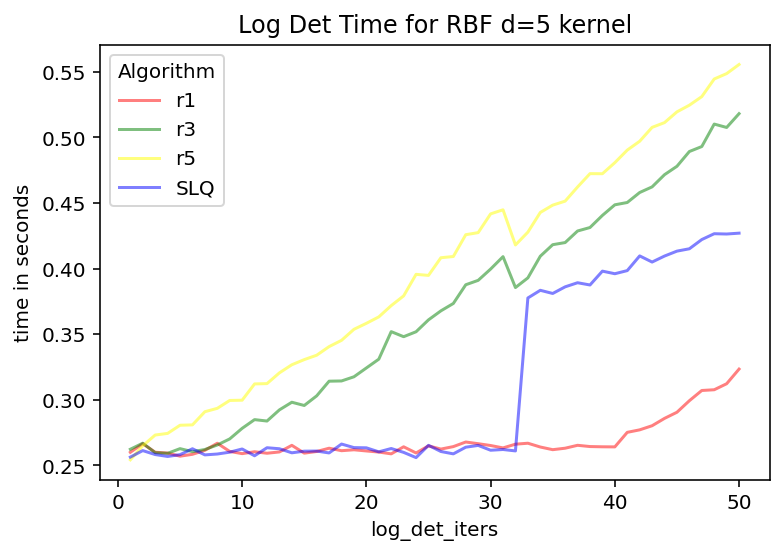}
	\includegraphics[width=6cm]{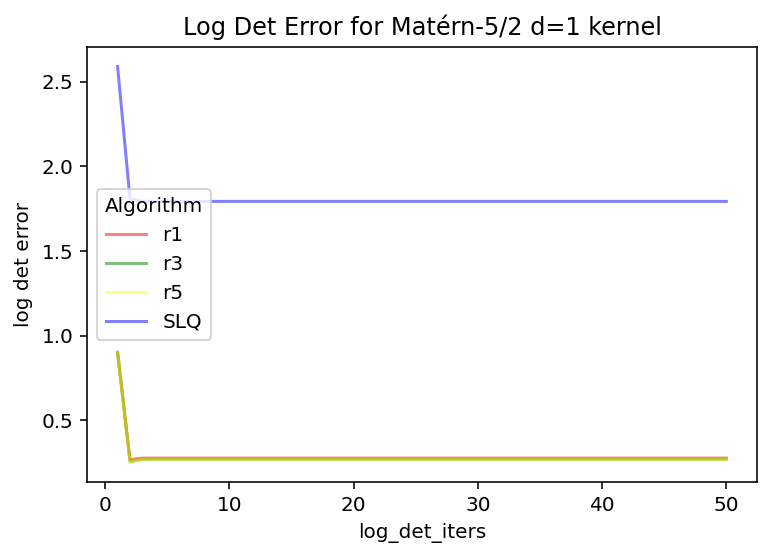}
	\includegraphics[width=6cm]{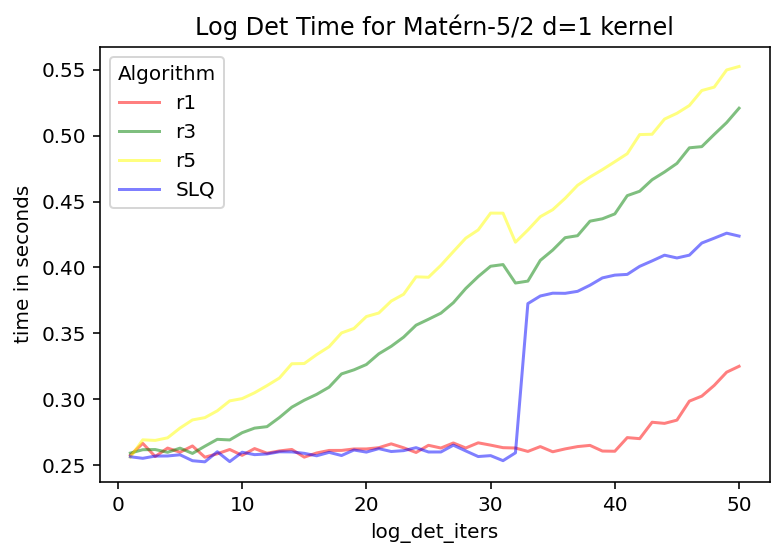}
	\includegraphics[width=6cm]{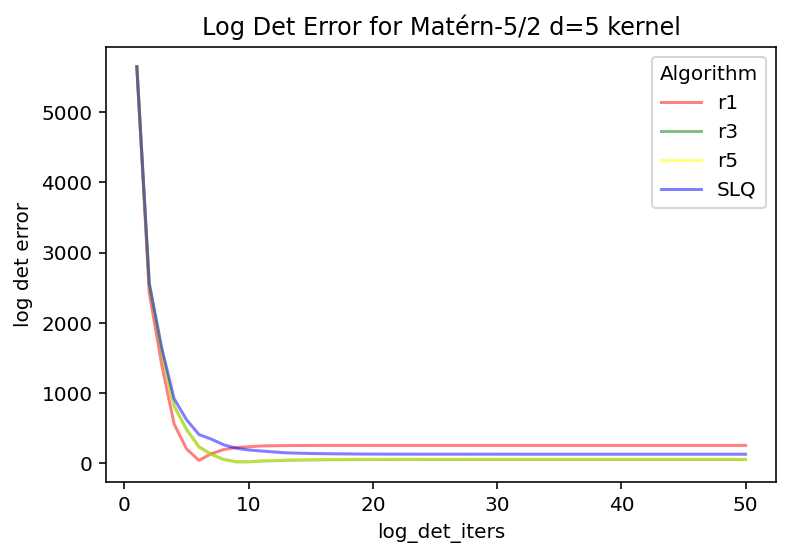}
	\includegraphics[width=6cm]{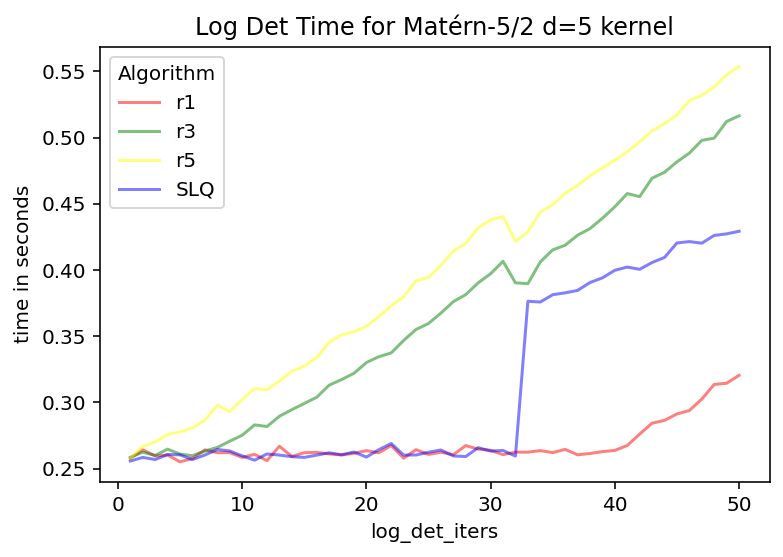}
	\end{center}
	\caption{Comparison of $\log \det$ algorithms when run with different values of the {\tt log\_det\_iters} parameter.  All measurements are averages over 100 randomly generated kernels with $n = 20,000$ as measured on a NVidia A100 GPU.}
	\label{log_det_iters_sweep}
\end{figure}

\end{document}